\documentclass[journal=nalefd,manuscript=article]{achemso}

\usepackage[version=3]{mhchem} 

\usepackage{gensymb}
\usepackage{graphicx}
\usepackage{dcolumn}
\usepackage{bm}
\usepackage{units}
\usepackage{upgreek}
\usepackage{makecell}
\usepackage{xcolor}

\mciteErrorOnUnknownfalse

\author{Roman M. Wyss}\altaffiliation{R. M. W. and M. P. contributed equally to this work.}
 \affiliation{Soft Materials
Department of Materials,
ETH Z\"urich, Z\"urich CH-8093, Switzerland}

\author{Markus Parzefall}\altaffiliation{R. M. W. and M. P. contributed equally to this work.}
\affiliation{
 Photonics Laboratory
ETH Z\"urich,
Z\"urich CH-8093, Switzerland}

\author{Cynthia M. Gruber}
\affiliation{
 Photonics Laboratory
ETH Z\"urich,
Z\"urich CH-8093, Switzerland}
\affiliation{IBM Research - Zurich, R\"uschlikon CH-8803, Switzerland}%

\author{Sebastian Busschaert}
\affiliation{
 Photonics Laboratory
ETH Z\"urich,
Z\"urich CH-8093, Switzerland}

\author{Karl-Philipp Schlichting}
\affiliation{Laboratory of Thermodynamics in Emerging Technologies
Department of Mechanical and Process Engineering,
ETH Z\"urich, Z\"urich CH-8092, Switzerland
}%

\author{Carin Rae Lightner}
\affiliation{Optical Materials Engineering Laboratory, 
Department of Mechanical and Process Engineering, 
ETH Z\"urich, 
Z\"urich CH-8092, Switzerland
}%

\author{Emanuel L\"ortscher}
\affiliation{IBM Research - Zurich, R\"uschlikon CH-8803, Switzerland}
 
\author{Lukas Novotny}
\affiliation{
 Photonics Laboratory
ETH Z\"urich,
Z\"urich CH-8093, Switzerland}
 
\author{Sebastian Heeg}
\email{sebastian.heeg@physik.hu-berlin.de}
\affiliation{
Photonics Laboratory
ETH Z\"urich,
Z\"urich CH-8093, Switzerland}
\affiliation{current affilitation: 
Department of Physics, Humboldt Universit\"at zu Berlin, 12489 Berlin, Germany}

\title{Freestanding and permeable nanoporous gold membranes for surface-enhanced Raman scattering}


\begin{document}



\begin{abstract}
Surface-enhanced Raman spectroscopy (SERS) demands reliable, high enhancement substrates in order to be used in different fields of application. Here, we introduce freestanding porous gold membranes (PAuM) as easy to produce, scalable, mechanically stable, and effective SERS substrates. We fabricate large-scale sub-\unit[30]{nm} thick PAuM, that form freestanding membranes with varying morphologies depending on the nominal gold thickness. These PAuM are mechanically stable for pressures up to $\unit[>3]{bar}$, and exhibit surface-enhanced Raman scattering with local enhancement factors of $10^4$ to $10^5$, which we demonstrate by wavelength-dependent and spatially resolved Raman measurements using graphene as a local Raman probe. Numerical simulations reveal that the  enhancement arises from individual, nanoscale pores in the membrane acting as optical slot antennas. Our PAuM are mechanically stable, provide robust SERS enhancement for excitation power densities up to \unit[10$^6$]{W\,cm$^{-2}$}, and may find use as a building block in flow-through  sensor applications based on SERS.
\end{abstract}



\section{Introduction}
Surface-enhanced Raman spectroscopy is a label-free, sensitive and compound-specific analytical technique, with an increasing number of applications in biomedical, environmental and surface sciences~\cite{RN956,Langer:2019a}. It relies on the chemical specificity of Raman spectroscopy in combination with the electromagnetic (EM) field enhancement in the vicinity of metallic nanostructures. The enhancement arises from localized surface plasmons (LSPs) that generate strong and localized near fields close to the metal. LSPs can be tuned in frequency by the nanostructures' material, shape and size~\cite{RN957}. Particularly strong enhancement arises from `hotspots' in the dielectric crevice between two or more metallic structures. Molecules and adsorbates located at these hotspot may experience enhancement up to a factor of $10^9$, enabling Raman spectroscopy down to the single molecule level~\cite{RN952,RN955}. Various methods for creating Au-based SERS substrates have been studied, including e-beam lithography \cite{RN964}, self assembly \cite{RN963, RN962} or templating \cite{RN958}, leading to a plethora of SERS structures such as densely packed gold nanopyramids \cite{RN967,RN965}, nanolits \cite{Chen:2009a} or nanostars~\cite{RN966}. Some of the drawbacks of current Au-based SERS substrates are the demanding manufacturing processes, limited scalability, and the laser intensity damage threshold of $\unit[\sim10^3 - 10^4]{W\,cm^{-2}}$, which limits SERS sensitivity~\cite{RN964,RN963,RN962,RN958,RN967,RN965,RN966,wasserroth2018graphene,zhu2014quantum,mueller2017evaluating,heeg2014plasmon,heeg2018probing}.

A particular intriguing application of SERS is flow trough sensing, where an analyte is detected by its Raman signature as it passes through a nanoslit that acts as a SERS hotspot. Current designs achieve single molecule sensitivity but require several processing steps including electron beam and UV lithography, which limits scalability and cost efficiency~\cite{Chen:2018a}. An attractive alternative to nanoslits are therefore SERS hotspots formed by nanopores that naturally arise during the formation or post-processing of metallic films. Such porous thin metal films have recently received attention in various fields such as plasmonics, photovoltaics, catalysis~\cite{zhang2014porous}, and electrochemistry~\cite{RN882,yu2006simultaneous}. SERS from supported porous thin films \cite{yan2016locally,novikov2016enhancement,Qian:2007a,Jiao:2011a} and dealloyed gold membranes have been studied, investigating their Raman enhancement as function of gap width \cite{qian2007surface,lang2011localized}. However, to the best of our knowledge, freestanding porous SERS structures with appreciable field enhancement, the ideal central building block of a SERS-based flow-through sensor, have not been studied thus far.

\begin{figure}[ht]
	\centering
	\includegraphics{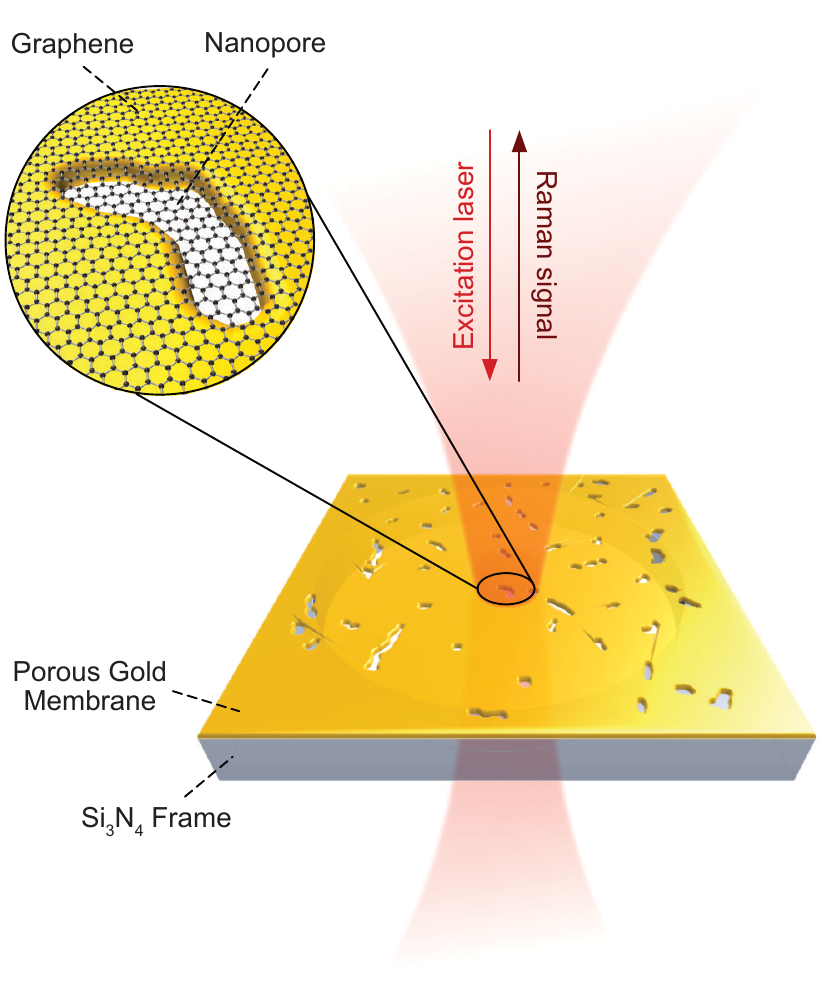}
	\caption{Illustration of a freestanding porous gold membrane that exhibits surface-enhanced Raman scattering. Graphene suspended over the nanopores probes and quantifies SERS enhancement.}
\label{fig0}
\end{figure} 

Here we demonstrate the fabrication and characterization of sub-\unit[30]{nm}  thick, freestanding porous metal membranes and establish them as efficient, permeable SERS substrates, cf. Fig.\ref{fig0}. Our porous metal films are manufactured on wafer scale without lithographic or advanced chemical assembly techniques. In contrast to previous work, our freestanding PAuM are intrinsically porous and mechanically stable up to $> \unit[3]{bar}$ of pressure, while being highly gas permeable. Using graphene placed over the membrane as a Raman probe, we observe local Raman enhancement of the order of $10^4$ to $10^5$ that is stable at unprecedented high excitation power densities $\sim \unit[10^6]{W\,cm^{-2}}$. Numerical simulations that take into account the nanopore geometry indicate that the enhancement arises from individual nanopores acting as hotspots for SERS. We show that the pores mimic plasmonic slot antennas---the electromagnetic complement to linear rod antennas. The porous gold membranes introduced in this work combine all relevant properties to act as highly effective and durable SERS substrates.  

\section*{Experiments and Results}

In the following we will elaborate on the fabrication of nanoporous gold membranes and analyze the pores' basic physical properties. Next, we will study their SERS performance in terms of signal enhancement, in conjunction with numerical simulations and damage threshold.

\subsection*{Membrane fabrication and characterization}
\begin{figure*}
	\includegraphics[width=0.95\textwidth]{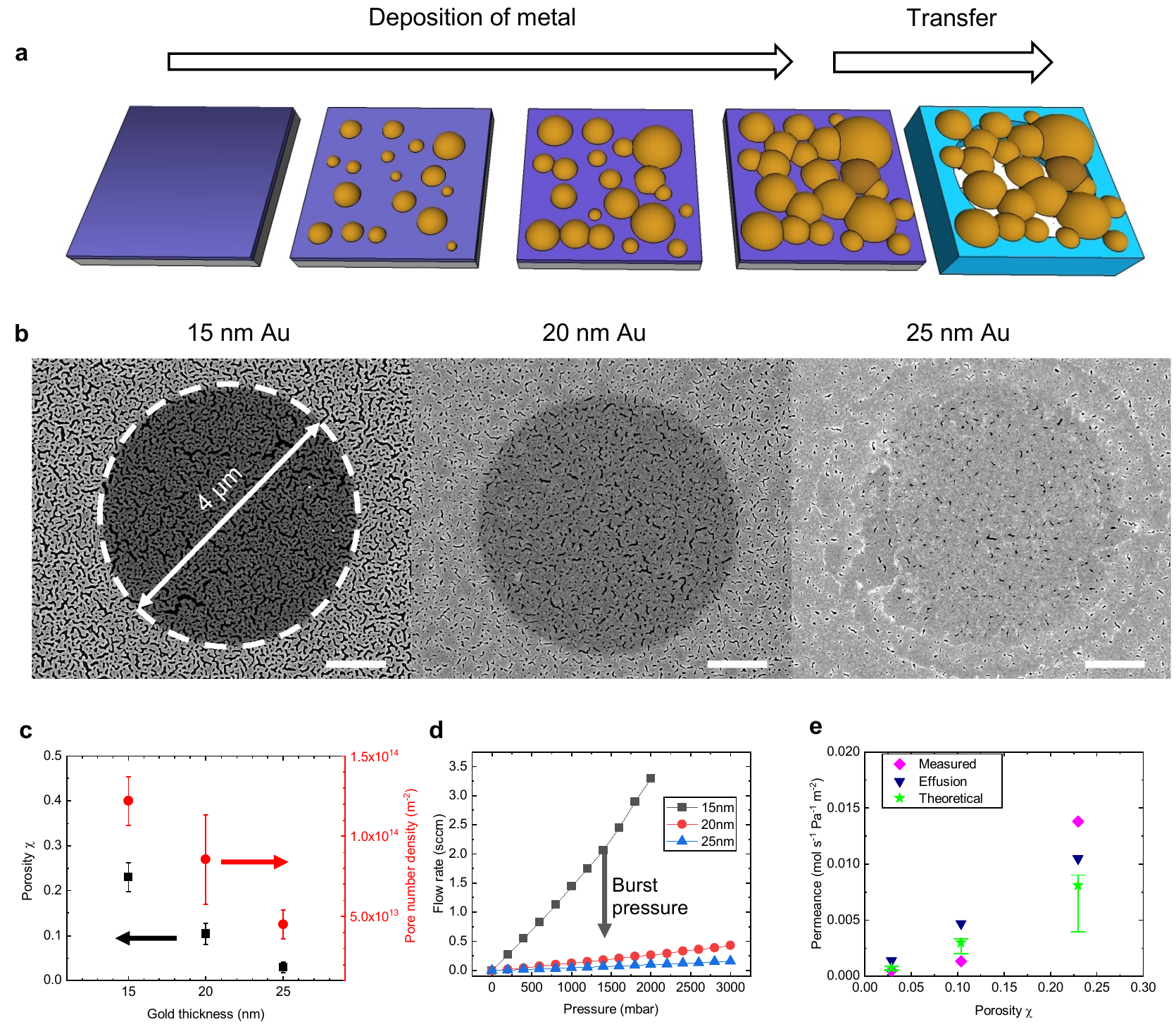}
	\caption{Porous Au membrane fabrication and characterization. (a) Schematic of the membrane manufacturing process by metal deposition via evaporation and subsequent transfer yielding freestanding porous gold membranes on Si$_3$N$_4$/Si chips. (b) Freestanding gold membranes with nominal thickness \unit[15]{nm}, \unit[20]{nm} and \unit[25]{nm} on a single hole in the Si$_3$N$_4$ chip. Scale bar: \unit[1]{$\upmu$m}. (c) Porosity and pore number density as a function of nominally deposited gold thickness.  (d) Gas flow versus applied pressure of Au membranes. A nonlinearity at \unit[1500]{mbar} indicates material failure of the \unit[15]{nm} membrane at this pressure (burst). (e) Measured and calculated permeances of porous gold membranes based on effusion and transmission probabilities. 
		\label{fig1}}
\end{figure*}

Freestanding gold membranes are produced by evaporation of Gold (Au) on Silicon/Silicon dioxide (Si/SiO$_2$) substrates and subsequent transfer to yield suspended membranes (Methods). During evaporation, the gold initially forms islands and clusters due to the different surface energies of Au (\unit[1.5]{Jm$^{-2}$}) and SiO$_2$ (\unit[0.3]{Jm$^{-2}$}) leading to nucleation and aggregation of gold clusters indicative of Volmer-Weber growth \cite{petrov2003microstructural,schwartzkopf2013atoms}. These islands grow in size and density, Fig.~\ref{fig1}(a), up to the point where they are large enough to touch, eventually forming a percolating network. Between the islands voids persist and give rise to a porous gold film below \unit[30]{nm} in thickness. While other nanoscopic gold structures have been reported to transit from ductile to brittle behavior \cite{li1992ductile,biener2005microscopic}, which prevents their use as freestanding structures, our nanoporous membranes easily span over micrometer-sized holes despite their small thickness.

Scanning electron microscope (SEM) images of freestanding PAuM with thicknesses ranging from \unit[15]{nm} to \unit[25]{nm} are shown in Fig.~\ref{fig1}(b) after transfer to a Si$_3$N$_4$/Si chip (see Methods), demonstrating that stable, freestanding membranes can be formed spanning over an $8 \times 8$ array of $\unit[4]{\upmu m}$ holes. We find that \unit[5]{nm} and \unit[10]{nm} PAuM suffer from frequent crack formation after transfer, so they cannot be used as membranes (Supporting Information S1). Our fabrication method is readily scaled to 4 inch wafer size, with uniform pores across the entire area (Supplementary Information S1). 

The pore density and porosity of the porous gold membranes are extracted from the SEM images in Fig.~\ref{fig1}(b) and are shown in Fig.~\ref{fig1}(c). A linearly decreasing porosity and pore number density is observed with increasing nominal Au thickness. The pores are circular and slit-like with typical radius/minor dimension of  \unit[33.5 $\pm$ 10.1]{nm} for \unit[15]{nm} PAuM and \unit[10.5 $\pm$ 5.2]{nm} for \unit[25]{nm} PAuM, while their length can be more than \unit[100]{nm}. The pore number density of $\sim$ \unit[10$^{14}$]{m$^{-2}$} is comparable to highly porous graphene membranes manufactured by self-assembly and lithographic techniques \cite{choi2018multifunctional}. Though the pore size, number, and  density/porosity cannot be controlled independently, highly porous membranes with controlled pore morphology are obtained without complex manufacturing.

We characterize the gas transport through the membranes by trans-membrane pressure driven nitrogen (N$_2$) flow (see Methods). The corresponding gas flow vs. pressure diagram is shown in Fig.~\ref{fig1}(d). We extract the permeance from the linear part (\unit[0 -- 1500]{mbar}) and plot it as a function of porosity in Fig.~\ref{fig1}(e). As expected, the permeance increases with increasing porosity. The gas transport through thin-film membranes is governed by effusive and collective transport \cite{RN95} as modelled in Supplementary Section S2.

The non-linearity in the flow \textit{versus} pressure graph of the \unit[15]{nm} PAuM in Fig.~\ref{fig1}(d) indicates mechanical failure (burst) of this membrane at $\sim \unit[1500]{mbar}$. The \unit[20]{nm} and \unit[25]{nm} membranes, on the other hand, sustain up to \unit[3]{bar} without breaking. Assuming a non-porous thin film with bulk material properties, we calculate the theoretical burst pressures BP$_{\rm TH}$ (bar) within film membrane theory (see Methods) and compare them to the measured burst pressures BP$_{\rm M}$ in table \ref{tab:pressure2}. At first sight, it might be surprising to find that $< \unit[30]{nm}$ membranes sustain more than \unit[3]{bar} of pressure, yet the burst pressure calculation yields values as high as \unit[5.24]{bar} for a \unit[25]{nm} PAuM, so that even thin film metal membranes can sustain reasonably high pressures. Taking into account, that the thin film theory predicts the burst behavior of non-porous structures, the measured and theoretical burst pressure are within reasonable agreement with each other. 

\begin{table}[h]
    \centering
    \begin{tabular}{l|c|c}
Membrane                    &    BP$_{\rm M}$ (bar)	 &   BP$_{\rm TH}$ (bar)\\
        \hline
    \unit[15]{nm}     & $1.5$ &  $3.14$ \\
    \unit[20]{nm}	  & $>3$  &  $4.19$  \\
    \unit[25]{nm}	  & $>3$  &  $5.24$ \\
 \end{tabular}
  \caption{Measured burst pressures BP$_{\rm M}$ and calculated burst pressures BP$_{\rm TH}$ for gold membranes of \unit[15]{nm}, \unit[20]{nm}, and \unit[25]{nm} thickness.}
    \label{tab:pressure2}
 \end{table}

Overall, the remarkable mechanical strength and porous metallic nature of our membranes suggest that they can be useful as easy-to-manufacture SERS substrates. 

\subsection*{Surface-enhanced Raman scattering}

\begin{figure*}
	\centering
	\includegraphics[width=\textwidth]{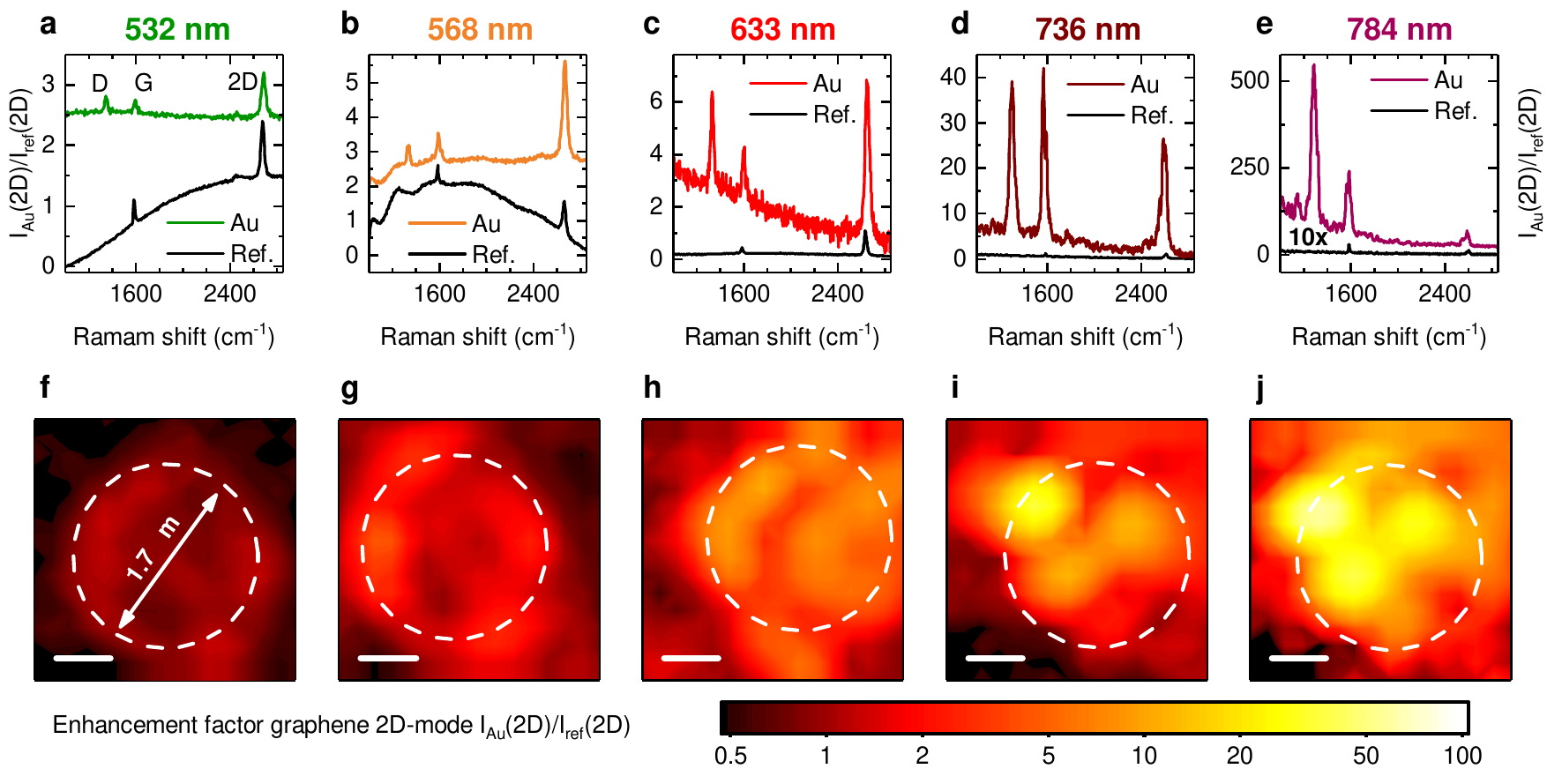}
	\caption{Wavelength-dependent SERS mapping of the nanopores in the freestanding gold membrane using graphene. Raman spectra of a graphene monolayer placed on top of a nanoporous gold membrane (thickness \unit[25]{nm}) for (a) \unit[532]{nm}, (b) \unit[568]{nm}, (c) \unit[633]{nm}, (d) \unit[736]{nm}, and (e) \unit[784]{nm} excitation wavelength. The intensity is normalized to the 2D mode of freestanding graphene (black) for reference. The SERS spectra in (a-e) are extracted at the locations of highest 2D intensities from Raman maps (f-j) recorded over the same circular freestanding membrane (dashed). Note that the enhancements factors displayed in (f) and (g) have been multiplied by factors of 10 and 5, respectively. Scale bar in (f-j) is \unit[500]{nm}.}
\label{fig3}
\end{figure*}

To develop a qualitative and quantitative understanding of SERS arising from the nanopores, we focus in the following on the \unit[25]{nm} thick membrane type, c.f. Fig.~\ref{fig1}(b). Its comparably low pore density allows us to study the enhancement arising from individual nanopores and how it is influenced by the pore geometry. Graphene is an ideal platform to probe SERS enhancement arising from plasmonic nanostructures ~\cite{RN965,RN959,schedin2010surface,kravets2012surface,heeg2013strained,wasserroth2018graphene}. A sheet of graphene can be easily interfaced with the nanopores by placing it on top of the PAuM, which we achieve here via a polymer-free transfer of CVD graphene, see Methods. The graphene conforms to the corrugated structure of the gold membrane and is suspended over the nanopores in the membrane as illustrated schematically in Fig.~\ref{fig0}. Graphene then serves as a two-dimensional detection material for the local SERS enhancement---reporting the average Raman enhancement from a horizontal plane of the pore---and allowing us to correlate the pores shape and size with the SERS enhancement. Furthermore, the Raman response of graphene is independent of the excitation wavelength and polarization \cite{RN45,ferrari2013raman}. For graphene placed on top of the gold membranes, any dependence of its Raman signal on these parameters can hence be attributed entirely to SERS enhancement arising from the nanopores~\cite{heeg2014plasmon,heeg2013strained,wasserroth2018graphene,schedin2010surface}. 
 
In Figs.~\ref{fig3}(a-e) we present the Raman spectra of graphene placed on a \unit[25]{nm}-PAuM for five different excitation wavelengths $\lambda_{\rm exc}$ between \unit[532]{nm} and \unit[784]{nm}. The spectra are extracted at the location of highest intensity from the Raman maps shown in Fig.~\ref{fig3}(f-j), which we recorded over the same freestanding membrane segment, see Methods. The dashed circles in Fig.~\ref{fig3}(f-j) mark the outline of the freestanding PAuM segment (diameter $\unit[1.7]{\upmu m}$, SEM image see Fig.~\ref{fig4}(a)). For all wavelengths, the intensity of the Raman spectra and Raman maps is normalized to the integrated 2D mode intensity of large area suspended graphene $\mathrm{I}_{\rm{ref}}$(2D), which we have prepared on a separate sample, see Supporting Information. Suspended graphene as a reference for SERS is beneficial because it resembles the configuration of the graphene suspended over the nanopores and is free of interference or other effects from an underlying substrate \cite{yoon2009interference}. For reference, the Raman spectra of freestanding graphene (black) are included in Figs.~\ref{fig3}(a-e). The broad background appearing in the Raman spectra of suspended graphene - in particular for \unit[532]{nm} and \unit[568]{nm} - arises from the partial overlap of the excitation with the Si$_3$N$_4$ frame below the focal plane of the objective, an effect that is otherwise suppressed by the gold membrane. 

Before we discuss SERS enhancement, it is useful to compare the pristine Raman response of graphene placed on the porous Au membrane to our suspended graphene reference. In both cases, the individual peak positions of the G- and the 2D-modes and their intensity ratio 2D/G observed in Fig.~\ref{fig3}(a) confirm the presence of single-layer graphene \cite{RN45,ferrari2013raman}. The drastic change in the 2D/G intensity ratio towards higher excitation wavelengths on the other hand is due to the reduced sensitivity of our Raman spectrometer at the wavelengths of the Raman scattered light (i.e. 2D mode at \unit[990]{nm} for $\lambda_{\rm exc}=\unit[785]{nm}$). Using suspended graphene as a reference, however, factors out the instrument sensitivity. Accordingly, we define the experimental SERS enhancement factors for the G- and 2D modes as $\mathrm{EF}_{\rm exp}=\mathrm{I}_{\rm{Au}}/\mathrm{I}_{\rm{ref}}$, and list the corresponding values extracted from the Raman spectra in Fig.~\ref{fig3}(a-e) in Table~\ref{tab:enhancement}. The D-mode just above $\unit[1300]{cm^{-1}}$ indicates the presence of defects in the CVD graphene on PAuM, while the (mechanically exfoliated) reference is largely free of defects. We therefore exclude the D-mode when evaluating SERS enhancement arising from the nanopores. 

In the following we discuss SERS enhancement from the nanopores, with a particular focus on its wavelength dependence and spatial distribution. For $\lambda=\unit[532]{nm}$, c.f. Fig.~\ref{fig3}(a) and (f), we do not observe any SERS enhancement, which is expected due to the d-band absorption of gold at this wavelength \cite{maier2007plasmonics,schedin2010surface,RN959}. Within the circular freestanding region, $\mathrm{I}_{\rm{Au}}\mathrm{(2D)/I}_{\rm ref}$(2D) ranges between $0.7$ and $1$. 
Outside the freestanding region, the Raman intensities of both the G- and the 2D-mode drop noticeably, an effect that we observe for all excitation wavelengths, c.f. Fig.~\ref{fig3}(f-j). One factor that contributes to this behavior is the high refractive index ($n \sim 2$) of the Si$_3$N$_4$ substrate which, compared to the suspended part of the membrane, directs a larger fraction of the Raman signal into the substrate instead of being back-scattered and collected~\cite{novotny2012principles}. 

\begin{table}
    \centering
    \begin{tabular}{c|c|c|c|c|c}
     $\lambda(\mathrm{nm})$& 532 & 568 & 633 & 736 & 784 \\
     \hline
     EF(G) & 0.9 & 3.0 & 13.3 & 169 & 202 \\
     EF(2D)  & 1.0& 4.3 & 10.5 & 42.7 & 60.0
    \end{tabular}
    \caption{Experimental SERS enhancement factors EF of the G- and 2D mode at locations of highest intensity in Fig.~\ref{fig3}(a-e) for different excitation wavelengths. The experimental enhancement factors at $\unit[784]{nm}$ translate to a local enhancement of $10^4$-$10^5$ at the nanopore, see simulationans and Fig.~\ref{fig4}.}
    \label{tab:enhancement}
\end{table}
For excitations at $\lambda_{\rm exc}=\unit[568]{nm}$, Fig.~\ref{fig3}(b,g), and $\lambda_{\rm exc}=\unit[633]{nm}$, Fig.~\ref{fig3}(c,h), we observe the onset and continuous increase of SERS enhancement over the entire freestanding region. We attribute this behaviour to longer nanopores with higher aspect ratios and overall area that come into resonance with the excitation, see Simulations. The maximal enhancement reaches $\mathrm{EF_{exp}(G)}=13.3$ and $\mathrm{EF_{exp}(2D)}=10.5$ at $\lambda_{\rm exc}=\unit[633]{nm}$ and remains uniform within a factor of two. It is not surprising to observe different SERS enhancement factors for the two Raman modes of graphene at the same location, as their comparably large energy separation ($\hbar\omega_{\rm G} \sim \unit[200]{meV}$, $\hbar \omega_{\rm 2D} \sim \unit[330]{meV}$) leads to variations in the spectral overlap of the Raman scattered light and a given plasmonic resonance, which in turn modifies the observed enhancement~\cite{RN959,wasserroth2018graphene}. 

The SERS enhancement changes drastically in spatial distribution and magnitude for excitation wavelengths above \unit[700]{nm}. We observe strong enhancement from three localized regions of the Au membrane, c.f. Figs.~\ref{fig3}(d,i) and (e,j). The peak enhancement extracted from Fig.~\ref{fig3}(e) reaches $\mathrm{EF}_{\rm G}=202$ and $\mathrm{EF}_{\rm 2D}=60$ for $\lambda_{\rm exc}=\unit[784]{nm}$. This suggests that the enhancement arises from very few or even single pores that act as SERS hotspots -- with local enhancements of $10^4-10^5$, see Simulations -- as compared the rather homogeneous enhancement observed for $\lambda<\unit[700]{nm}$, where many pores contribute to the overall enhancement. Importantly, these hotspots are characteristic of the entire Au-membrane and not limited to the freestanding circular region shown Fig.~\ref{fig3}. We observed the same behavior for two additional circular freestanding regions for which wavelength-dependent Raman mapping was performed, see Supporting Information. 

\subsection*{Simulations}

\begin{figure*}
	\centering
	\includegraphics{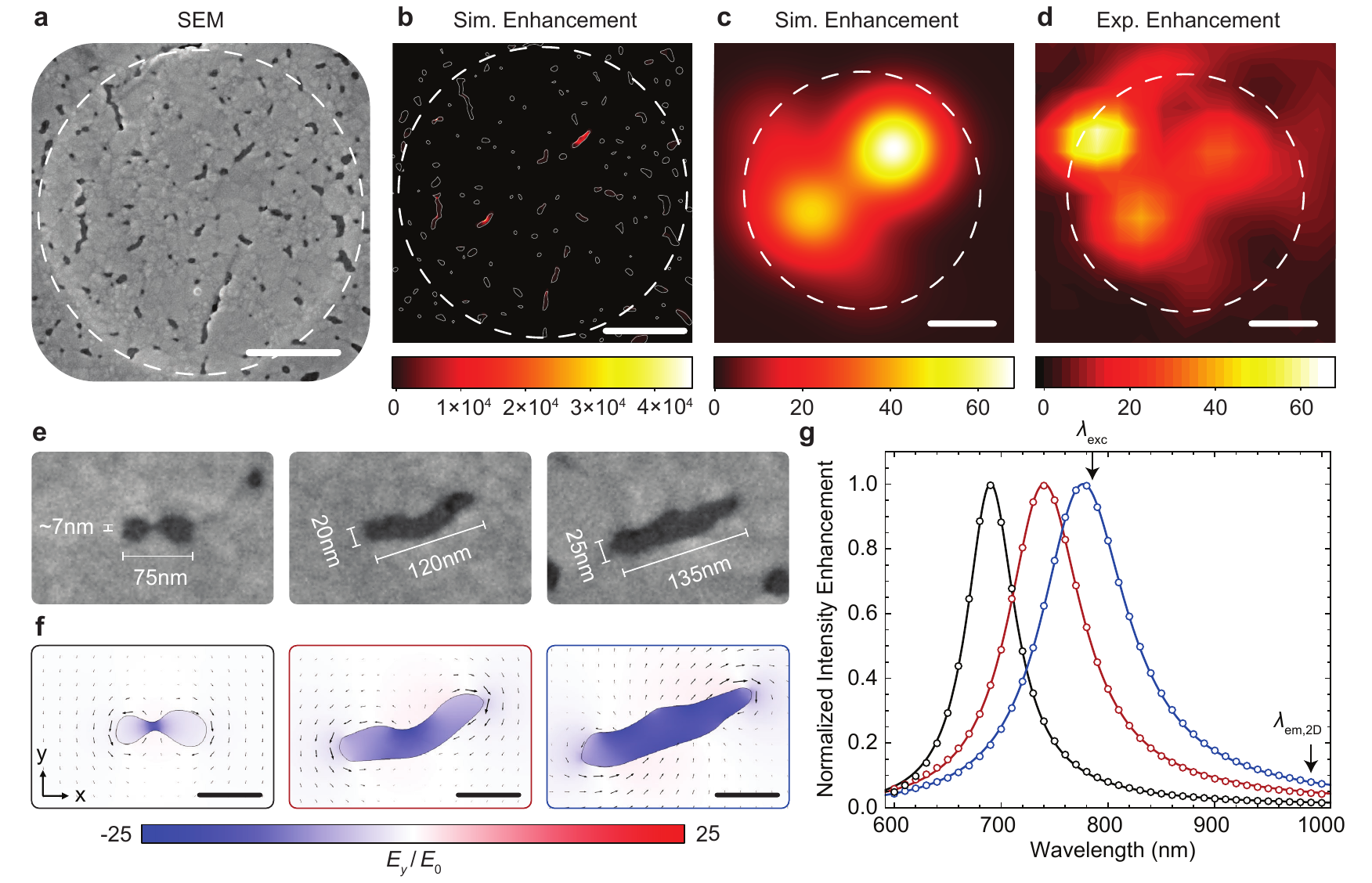}
	\caption{Numerical analysis of porous gold membranes. (a) SEM image of a freestanding Au membrane (\unit[25]{nm} nominal thickness), covered by monolayer graphene. (b)~Simulated local enhancement $\mathrm{EF}_{\rm loc}$ (cf. Eqn.~(\ref{eq:EF})) of the 2D Raman mode intensity for an excitation wavelength $\lambda_{\rm exc}$ of \unit[784]{nm}, overlaid with the outlines extracted from (a). (c) As (b), convoluted with a Gaussian function, the full width at half maximum (FWHM) of which is given by $\lambda_{\rm exc} / (2 \mathrm{NA})$, where $\mathrm{NA}=0.8$ is the numerical aperture of the objective. (d)~Experimental map of the enhancement of the integrated 2D intensity at $\lambda_{\rm exc} = \unit[784]{nm}$, reproducing the data shown in Fig.~\ref{fig3}(e). (e)~Magnified SEM images of three nanopores. (f)~Simulated enhancement of the $y$-component of the electric field for the three nanopores shown in (e) under $y$-polarized plane wave illumination for $\lambda_{\rm exc} = \unit[784]{nm}$. Black arrows indicate magnitude and direction of the current density. (g)~Simulated, normalized intensity enhancement spectra in the center of the three nanopores (empty circles), fitted by Lorentzian functions (lines). Colors correspond to the colored frames in (f). The dashed outlines in (a-d) mark the position of the suspended membrane. Scale bar in (a-d) is \unit[500]{nm}. Scale bar in (f) is \unit[50]{nm}.
		\label{fig4}}
\end{figure*}

To gain insight into the origin of the observed Raman enhancement and the emergence of hotspots localized within the freestanding Au membrane towards longer excitation wavelengths, we perform numerical electrodynamics simulations of the geometry based on high-resolution scanning-electron microscope (SEM) images, as shown in Fig.~\ref{fig4}(a) for the membrane studied in Fig.~\ref{fig3}. Detailed information regarding the simulations is given in the Methods section.

In brief, we extract a two-dimensional map of outlines of the nanopores within and around the suspended Au membrane from the SEM image shown in Fig.~\ref{fig4}(a). We then extrude this map to the nominal thickness of the membrane (\unit[25]{nm}) and simulate the enhancement of the electric field amplitude $|\mathbf{E}|/|\mathbf{E_0}|$, where $|\mathbf{E}_0|$ is the amplitude of the plane wave excitation. The local Raman enhancement factor EF$_{\rm loc}$ is then given as~\cite{RN956,maier2007plasmonics}
\begin{equation}\label{eq:EF}
    \mathrm{EF}_{\rm loc} = \left( \frac{\left| \mathbf{E} \left( \lambda_{\rm exc} \right) \right| }{\left| \mathbf{E_0} \left( \lambda_{\rm exc} \right) \right| } \right)^2 \times \left( \frac{\left| \mathbf{E} \left( \lambda_{\rm em} \right) \right| }{\left| \mathbf{E_0} \left( \lambda_{\rm em} \right) \right| } \right)^2 ,
\end{equation}
where $\lambda_{\rm em}$ is the emission wavelength of the Raman mode. 

Figure \ref{fig4}(b) shows the simulation results of Eqn.~(\ref{eq:EF}) for a cross section through the center of the membrane. The simulations were carried out for $\lambda_{\rm exc} = \unit[784]{nm}$ and $\lambda_{\rm em,2D} = \unit[990]{nm}$, corresponding to the 2D Raman mode of graphene. The polarization of the exciting plane waves corresponds to the experimental polarization (cf. Methods). The simulated enhancement map is overlaid with the outlines of the nanopores as extracted from the SEM image in Figure \ref{fig4}(a), suggesting that, the enhancement is dominated by a few selected nanopores, in agreement with the experimental observation of localized hotspots. The simulation further suggests that--- within those nanopores---the Raman enhancement locally reaches values of $10^4$ to $10^5$. 
To allow for a direct comparison with our experiments, we further convolute the enhancement map with a two-dimensional Gaussian function, emulating the diffraction-limited nature of the confocal measurement. The results are shown in Fig.~\ref{fig4}(c). The spatial distribution of enhancement is in qualitatively good agreement with the corresponding experimentally obtained map displayed in Fig.~\ref{fig4}(d). Furthermore, both experiment and simulation are in agreement regarding the absolute magnitude of the Raman enhancement. We find similarly good agreement between experiment and simulation for two additional freestanding sections of the membrane, see Supplementary Information. 

Next, we carry out a detailed analysis of the three nanopores which exhibit the highest enhancement factors. Magnified SEM images of these nanopores are shown in Fig.~\ref{fig4}(e). All three are similarly oriented and elongated along the same axis. Additionally, the first pore is narrowed down to a minimum width of $\sim \unit[7]{nm}$. Figure~\ref{fig4}(f) further demonstrates that all three nanopores display the same electric field and current distributions, suggesting that the nanopores support localized resonances that are determined by their size and shape. These resonances are indeed observed when analyzing the intensity enhancement in the center of each nanopore as a function of wavelength, shown in Fig.~\ref{fig4}(g). The spectral resonance positions of the two larger pores agree very well with the evolution of enhancement as a function of wavelength observed in Fig.~\ref{fig3}(a-e). Our simulations do not reproduce the experimentally observed trend for the smallest pore. This, however, is not surprising as our simulated geometry neglects the surface morphology of the pores for reasons of computational efficiency. The morphological details have the strongest effect on resonance position and enhancement for pores that are narrower than the nominal film thickness, as is the case here. This explains the mismatch between experiment and simulation with regards to the smallest pore. 

Our analysis shows that nanopores act as slot antennas---the electromagnetic complement to linear rod antennas \cite{hentschel2013babinet,yang2014accessing}---exhibiting shape and size-dependent LSP resonances (cf. Supplementary Information) . These resonances are accompanied by strongly enhanced near-fields, leading to the observed Raman intensity enhancement. Slot antennas are most efficiently excited with the electric field polarized parallel to their minor axis \cite{gordon04strong,garcia2010light}, in agreement with the relative orientation of the three nanopores shown in Fig.~\ref{fig4}(e,f) with respect to the incident plane wave polarization. The resonance position of slot antennas shifts towards longer wavelengths with increased aspect ratio \cite{koerkamp2004strong,garcia2005transmission}. In addition we find that slot antennas exhibit a stronger enhancement at higher aspect ratios, as well as when they are suspended (cf. Supplementary Information). This explains both the overall increase in enhancement towards longer wavelengths as well as the localization of the highest enhancement to the suspended sections of the Au membrane, as observed in Fig.~\ref{fig3}(a-e). 

Finally, our simulations of slot antennas suggest that the enhancement is fairly constant across the entire depth of the nanopores (cf. Supplementary Information). Hence, the enhancement probed by the graphene membrane is a faithful representation of the enhancement present in the pore’s entire volume, which is the relevant figure of merit for freestanding flow-through SERS structures. 

\subsection*{Stable SERS under high illumination powers}

\begin{figure}
	\centering
	\includegraphics{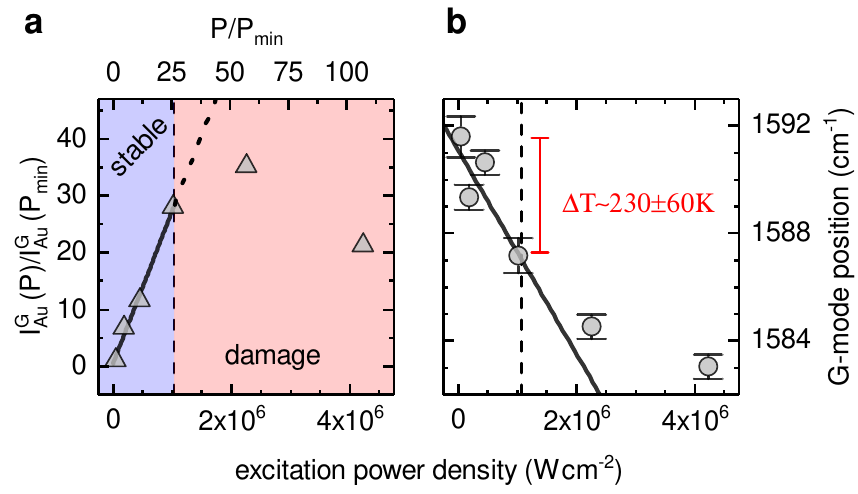}
	\caption{SERS stability under high illumination powers. (a) G-mode intensity and (b) position of graphene at a nanopore hotspot for various excitation power densities. The hotspot is stable up to $\mathrm{P}\sim\unit[1.1\times10^6]{\mathrm{W}\mathrm{cm}^{-2}}$ for which the G-mode downshift suggest a local temperature increase $\Delta T\sim 230K$.}\label{fig5}
\end{figure}

The sensitivity of a SERS based sensor---the intended application of the porous gold membranes presented in this work---does not depend on the local Raman enhancement at a hotspot itself, but rather on the absolute measured SERS signal. The latter scales linearly with the excitation power. The more excitation power a SERS substrate can sustain, the better it performs as a sensor. To evaluate the stability of our \unit[25]{nm}-PAuM as a SERS substrate, we recorded graphene Raman spectra at a nanopore hotspot comparable to that described in Fig.~\ref{fig3}(e,j) and Fig.~\ref{fig4} at $\lambda=\unit[784]{nm}$ for different excitation power densities P (see Methods). The graphene G-mode intensity $\mathrm{I}^{\rm G}_{\rm Au}(\mathrm{P})$ at the hotspot as a function of power density is shown in Fig.~\ref{fig5}(a). Here $\mathrm{I}^{\rm G}_{\rm Au}(\mathrm{P})$ is normalized to the intensity $\mathrm{I}^{\rm G}_{\rm Au}\mathrm{(P)}$ recorded at the lowest power density $\mathrm{P_{\rm min}}\sim \unit[4 \times 10^4]{\mathrm{W\,cm}^{-2}}$, which corresponds to a power of $\unit[350]{\upmu W}$ in the laser focus of our setup. 

Figure~\ref{fig5}(a) shows that the enhancement of the G-mode increases linearly (solid line) with the excitation power density as $\mathrm{I}^{\rm G}_{\rm Au}(\mathrm{P})/\mathrm{I}^{\rm G}_{\rm Au}(\mathrm{P_{\rm min}})=1.06\pm0.07\,\mathrm{P}/\mathrm{P}_{\rm min}$ between $\mathrm{P}_{\rm min}\sim\unit[4\times10^4]{\mathrm{W}\mathrm{cm}^{-2}}$ and $\mathrm{P}\sim\unit[1.1\times10^6]{\mathrm{W}\mathrm{cm}^{-2}}$. This behavior indicates that both the nanopore forming the SERS hotspot and the suspended graphene membrane remain functional and deliver reliable SERS enhancement up to $\mathrm{P}\sim\unit[1.1\times10^6]{\mathrm{W}\mathrm{cm}^{-2}}$, which we confirm by cycling back to $\mathrm{P}_{\rm min}$ after every increase in P without loss in G-mode intensity, see Supporting Information.  For larger power densities, $\mathrm{I}^{\rm G}_{\rm Au}(\mathrm{P})$ increases further without following the linear trend (dashed line). This relative decrease in SERS enhancement indicates the onset of morphological changes within the nanopore and/or damage of the graphene membrane. At $\mathrm{P_{\rm max}}\sim\unit[4.2\times10^6]{\mathrm{W}\mathrm{cm}^2}$, the nanopore hotspot with its graphene Raman probe eventually fails irreversibly.

To gain insight into the failure mechanism of the SERS hotspot, we extract the local temperature from the change of the G-mode position of graphene with increasing laser power, Fig.~\ref{fig5}(b). For  $\mathrm{P}=\unit[1.1 \times 10^6]{\mathrm{W}\mathrm{cm}^{-2}}$, the highest power with stable SERS enhancement, the G-mode is downshifted by $\unit[3.6\pm0.9]{cm^{-1}}$ with respect to P$_{\rm min}$, which corresponds to a temperature increase of $\unit[230\pm60]{K}$ (see Methods) \cite{calizo2007temperature}. Assuming no significant heating for P$_{\rm min}$ and thermal equilibrium between the suspended graphene and the Au membrane, we infer a local hotspot temperature of $\unit[\sim525\pm60]{K}$, low compared to the melting point of bulk gold ($\unit[\sim1330]{K}$), nanometer-scale gold nanoparticles ($\unit[\geq 950]{K}$)~\cite{koga2004size}, as well as the breakdown temperature of graphene in air ($\unit[875]{K}$)~\cite{Dorgan:2013a}. For $\mathrm{P}=\unit[2.2 \times 10^6]{\mathrm{W}\mathrm{cm}^{-2}}$, the linear trend in Fig.~\ref{fig5}(b) suggests that the nanopore-graphene system fails at a local temperature of around $\unit[855]{K}$. A temperature-induced breakdown of the suspended graphene Raman probe is therefore rather unlikely. It should be noted, however, that the diffusion of gold atoms is non-negligible at room temperature. We, therefore, hypothesize that the most likely failure mechanism is the disintegration of the hotspots due to the accelerated gold diffusion and the simultaneous reshaping of the membrane. 

\section*{Discussion}

Compared to alternative SERS substrates, our porous Au membranes show an outstanding durability with respect to illumination power. The membrane acts as its own heat sink thanks to the superior heat conduction inherent to metals and the corresponding efficient heat dissipation away from the illuminated area. Stable SERS enhancement for $\mathrm{P}=\unit[1.1 \times 10^6]{\mathrm{W}\mathrm{cm}^{-2}}$ exceeds the limits reported in the literature for conventional self-assembled ($\unit[\sim 10^3]{W\,cm^{-2}}$) and lithographically fabricated metallic SERS nanostructures ($\unit[\sim10^3 - 10^4]{W\,cm^{-2}}$) by up to three orders of magnitude \cite{RN964,RN963,RN962,RN958,wasserroth2018graphene,zhu2014quantum,mueller2017evaluating,heeg2014plasmon,heeg2018probing}, see Supporting information for an overview, and demonstrates the extraordinary thermal stability of nanoporous Au membranes introduced in this work. 

The $10^2 - 10^3$-fold increase in excitation power that can be sustained by our Au-membranes as compared to alternative SERS substrates corresponds directly to a $10^2 - 10^3$-fold increase in total SERS signal and hence SERS sensitivity. This compensates for the weaker maximum local SERS enhancement of the nanopores $(\sim 10^5)$ compared to that reported for hotspots of supported Au dimers with sub-nm gaps  (up to $\sim 10^8-10^9)$~\cite{zhu2014quantum}. The nanopores, on the other hand, provide enhancement over far greater volumes than sub-nm hotspots, which makes them particularly suitable as permeable SERS sensor. 

In this work, we chose to focus on the \unit[25]{nm} thick membrane with the lowest available pore density, as this allowed us to assess the performance of PAuM based on isolated pores that act as SERS hotspots. Large scale Raman mapping of the \unit[20]{nm} thick PAuM (see Supporting information) shows that the total SERS enhancement increases approximately linearly with the pore number density, c.f. Fig~\ref{fig1}, and becomes more spatially homogeneous. The optimal PAuM thickness for a particular SERS experiment or application will depend on the specific needs with respect to permeance, mechanical stability, hotspot density and overall SERS enhancement.  

\section*{Conclusions}

In conclusion, we have introduced freestanding nanoporous gold membranes as a simple, scalable, novel and effective SERS platform. Our suspended gold films of \unit[15]{nm} - \unit[25]{nm} thickness are highly porous and mechanically stable up to $>$ \unit[3]{bar} of pressure, allowing for their use as on-surface as well as permeable SERS substrates. Using graphene as a Raman probe in combination with excitation energy dependent Raman mapping, we observe a $10^5$ SERS enhancement from hotspots formed by individual nanopores, in agreement with numerical simulations. A general benefit of freestanding SERS structures is an increased emission of the Raman scattered light into the detection half-space compared to supported structures. Our Au membranes withstand excitation powers up to $10^6$W\,cm$^{-2}$ – adding two orders of magnitude to their sensitivity as compared to conventional SERS nanostructures – with moderate temperatures at the nanopore hotspots. Optimizing the fabrication parameters towards nanopores with a preferential size, shape and orientation, i.e. by prepatterning the substrate for membranes formation, has the potential to achieve even higher and more homogeneous enhancement.

\section*{Methods}

\subsection*{Manufacturing and Transfer of porous Ai membranes} 
Au membranes are produced by e-beam evaporation of Au (Evatec, deposition rate $\unit[0.2]{nm/s}$, distance from source: \unit[600]{mm}) on Silicon (Si) wafers with \unit[30]{nm} thick thermally grown silicon dioxide (SiO$_2$). The wafers have been cleaned by acetone-isopropyl alcohol rinsing followed by \unit[1]{min} of oxygen plasma (600W, O$_2$ flow \unit[25]{sccm}, TePla).
The nominal thickness of the PAuM membranes manufactured here are \unit[5]{nm}, \unit[10]{nm}, \unit[15]{nm}, \unit[20]{nm} and \unit[25]{nm} as indicated by the quartz crystal.

The porous gold films are transferred similarly to 2D materials, such as graphene, to yield freestanding membranes \cite{li2009large,reina2008transferring}. A \unit[120]{nm} thick polymethylmethacrylate (PMMA, in Anisol $\unit[2]{w\%}$) protection layer is coated and subsequently floating etched in buffered hydrofluoric acid (BHF). Under-etching leads to release of the Au/PMMA from the Si wafer resulting in a floating film, which is rinsed $\unit[2\times20]{min}$ in DI-water and transferred to a Si/Si$_3$N$_4$ substrate with arrays of holes forming freestanding PAuM. 
 
Chips with holes of \unit[4]{$\upmu$m} in arrays of $8 \times 8$ are manufactured following the procedure from ref \cite{RN95}. In brief, wafers (from Si-Mat, 150nm Si$_3$N$_4$ on Si, ultra low stress) are photolithographically defined and etched by fluor-based chemistry using reactive ion etching (RIE) and subsequent KOH etching to form freestanding Si$_3$N$_4$ membranes. Into these freestanding Si$_3$N$_4$ membranes, the holes are defined lithographically and etched by RIE.

\subsection*{Burst pressure calculation based on thin film theory}
For a thin membrane on a circular support of diameter D we can assume that the membrane is only loaded by in-plane stress, where the burst pressure is defined by
\begin{equation}\label{eq:burstpressure}
\Delta P =\frac{t}{D}\sqrt{\frac{96(1.026-0.79\nu-0.233\nu^2)\sigma^3}{E}},
\end{equation}
where $E=\unit[79]{GPa}$ is the Young’s modulus, $\nu=0.4$ is the Poisson ratio, $\sigma=\unit[220]{MPa}$  is the breaking strength, $t$ is the membrane thickness and $D=\unit[4 \times 10^{-6}]{m}$ is the diameter of the support hole. In this calculation, the porosity of the gold membrane is neglected leading to an upper bound value for the burst pressure.

\subsection*{Support-free transfer of graphene to porous gold}
Graphene was grown in a cold-wall chemical vapor deposition (Aixtron ltd.) as described in Ref.~\cite{RN95} yielding predominately monolayer graphene on copper (Alfa Aesar 46986). To avoid contamination of graphene by polymer residues, graphene was transferred polymer-free, following the method reported in Ref.~\cite{zhang2016versatile}. The as-grown graphene on copper is placed at the interface of copper etchant (0.5M ammonium persolfate (APS)) and hexane. Once the copper has dissolved entirely, the graphene is removed from the etchant and placed at the interface of DI-water hexane, and rinsed for \unit[30]{min} before being scooped with the target substrate and dried. 

\subsection*{Raman spectroscopy} Wavelength dependent Raman mapping was performed on a home-built Raman setup equipped with a xy-Piezo stage (MadCity Labs) in backscattering configuration using a $100\times$ (NA $0.9$) objective. A Coherent Innova $70$C Ar/Kr laser (\unit[532]{nm}, \unit[568]{nm}), a Thorlabs HeNe laser \unit[633]{nm}) and a wavelength tunable Ti:Sa laser (\unit[736]{nm}, \unit[784]{nm}) served as linearly polarized excitation sources. For all excitation wavelengths, the laser power was kept below \unit[1.5]{mW} for graphene on the PAuM and below \unit[0.5]{mW} for the suspended graphene reference. For each wavelength, the reference was measured directly after taking a Raman map without changing any instrument parameters. Integration times were up to $\unit[10]{\rm s}$ for the maps and up to $\unit[90]{s}$ for the reference. The Raman maps were acquired with $\unit[150]{nm}$ step-size. The polarization in all Raman measurements was set to $\unit[22.5]{\degree}$ from the vertical axis in Fig.~\ref{fig3}.  

Power dependent measurements were performed on a Horiba LabRam HR Evolution system equipped with a \unit[785]{nm} Toptica XTRA II high power single frequency diode laser and a  LMPLFLN100X Olympus objective (NA 0.9). The laser power was measured after the objective mount but without the objective. Taking into account the transmission efficiency of our objective ($75\%$ at \unit[785]{nm}) and the overfilling of the objective's back aperture, the effective power in the laser focus given in the main paper amounts to $\unit[50]{\%}$ of the measured power.

The local temperature T can be extracted from the graphene G-mode position $\omega$ using the relation
$\mathrm{T}=(\omega-\omega_0)\chi^{-1}$,
where $\omega_0$ is the room-temperature peak position and $\chi=\unit[-1.62\times10^{-2}]{cm^{-1}K^{-1}}$ is the temperature coefficient describing the G-mode's shift in position per $\unit[1]{K}$ change in temperature~\cite{calizo2007temperature}.

\subsection*{Scanning electron microscopy and image analysis} SEM was performed using a dual beam (FEI Helios) and a single beam (FEI Magellan) system at \unit[5]{kV} acceleration voltage and \unit[50]{pA} current. The pore dimensions were extracted from the SEM images using the software ImageJ. 

\subsection*{Gas flow measurements} The permeance was characterized using a self-built setup described in detail elsewhere \cite{RN95}. In short, the membrane is placed in a fixture with the gold facing upstream. A mass flow controller (MFC MKS) is used to increase the pressure on the feed side of the membrane, observed by a manometer. Downstream, a mass flow meter (MFM MKS) is used to record the flow rate, which is converted to molar permeance by normalizing with the applied pressure and the total freestanding area of the membrane ($\sim$ \unit[800]{$\mu$m$^2$}). 

\subsection*{Simulations}

Numerical finite element simulations were carried out using the wave optics module within the commercial software package COMSOL Multiphysics~5.3a. Outlines of the nanopores across a $\unit[1.8 \times 1.8]{\upmu m^2}$ large region of the Au membranes are extracted from high resolution SEM images, including the circular suspended region which is $\unit[1.7]{\upmu m}$ in diameter. These outlines are further extruded to the nominal height of \unit[25]{nm} of the membrane. Both half-spaces below and above as well as the pores within the membrane are filled with vacuum ($n=1$). Empirical values are used for the dielectric function of gold \cite{johnson1972optical}. The field enhancement $\left| \mathbf{E} \left( \lambda \right) \right| / \left| \mathbf{E_0} \left( \lambda \right) \right| $ at a given wavelength $\lambda$ is determined in two steps. First, we simulate the fields for a planar gold film without nanopores under plane wave illumination. Second, we determine the altered fields in the presence of nanopores. 

\subsection*{Data Availability}

The data that support the findings of this study are available from the corresponding author upon reasonable request.

\suppinfo

S1 - Characterization of freestanding Au membranes with \unit[5-25]{nm} thickness; S2 - Gas transport through Au Membranes; S3 - Freestanding graphene as Raman reference; S4 - SERS of two additional graphene covered, suspended Au-membranes. S5 - Numerical analysis of regular slot antennas in a thin Au film. S6 - SERS stability under high illumination powers. S7 - Large-scale SERS mapping of 20 nm thick porous Au membrane.

\section*{Acknowledgements}
R.M.W. acknowledges the Binnig and Rohner Nanotechnology Center. S.H. acknowledges financial support by ETH Z\"urich Career Seed Grant SEED-16 17-1 and by the German Research Foundation’s (DFG) Emmy Noether Programme (project
number 433878606). This research was partially funded by Swiss National Science Foundation (grant no. 200020$\_$192362/1). The authors thank N.Clark for providing the suspended graphene reference sample and acknowledge the use of the facilities at the Scientific Center for Optical and Electron Microscopy (ScopeM) at ETH Z\"urich.

\section*{Author contributions}
S.H. and R.M.W. conceived the project. R.M.W. manufactured and characterized the gold membranes. K.P.S. synthesized and transferred graphene. S.H. performed and evaluated the Raman measurements. S.B. performed SEM imaging. C.G., C.L. and E.L. contributed to the initial Raman characterization. M.P. performed all numerical simulations. R.M.W, M.P. and S.H. jointly interpreted the data, discussed the results and wrote the manuscript. All authors participated in the interpretation of the data and contributed to the final manuscript. R.M.W. and M.P. contributed equally to this work. S.H. coordinated and supervised the project.

\section*{Competing financial interests}
R.M.W. holds a patent for the manufacturing process of porous gold membranes ( EP18207023.5). The authors declare no other financial interests. 

\clearpage

\bibliography{Bib_porousgold}

\newpage

\section*{Supporting Information}

\section{Characterization of freestanding Au membranes with  \unit[5]{nm} - \unit[25]{nm} thickness}

\begin{figure*}[hp]
	\includegraphics[width=0.75\textwidth]{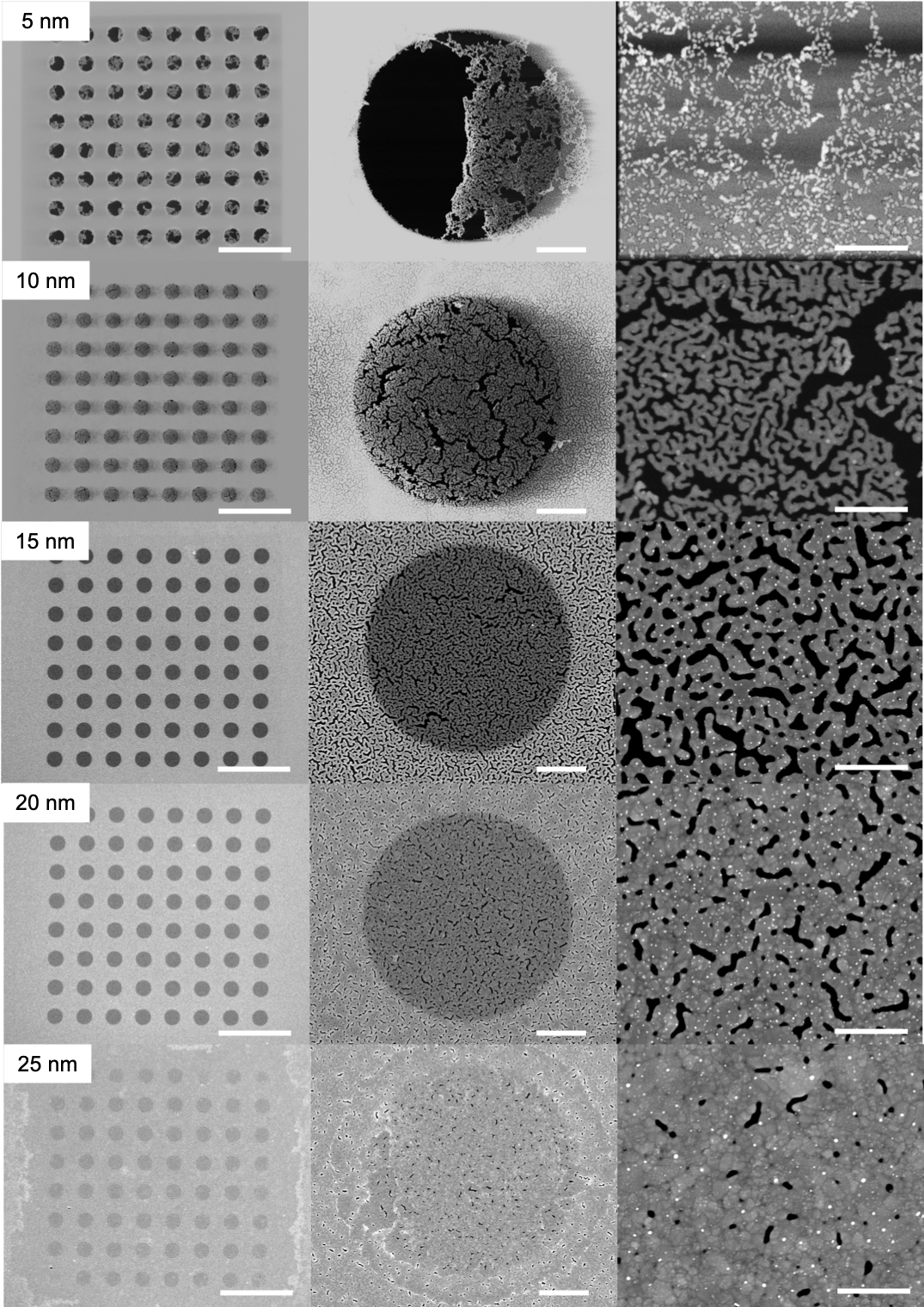}
	\caption{Full results of freestanding Au membranes from \unit[5]{nm} (top) to \unit[25]{nm} (bottom) in thickness, scale bars from left to right: \unit[20]{$\mu$m},\unit[1]{$\mu$m} and \unit[300]{nm}. The \unit[5]{nm} and \unit[10]{nm} thick films cannot form a suspended membrane or exhibit cracks and fractures, so they are omitted in the subsequent investigation.
	\label{fullresults}}
\end{figure*}

\begin{figure*}[hp]
	\includegraphics[width=0.85\textwidth]{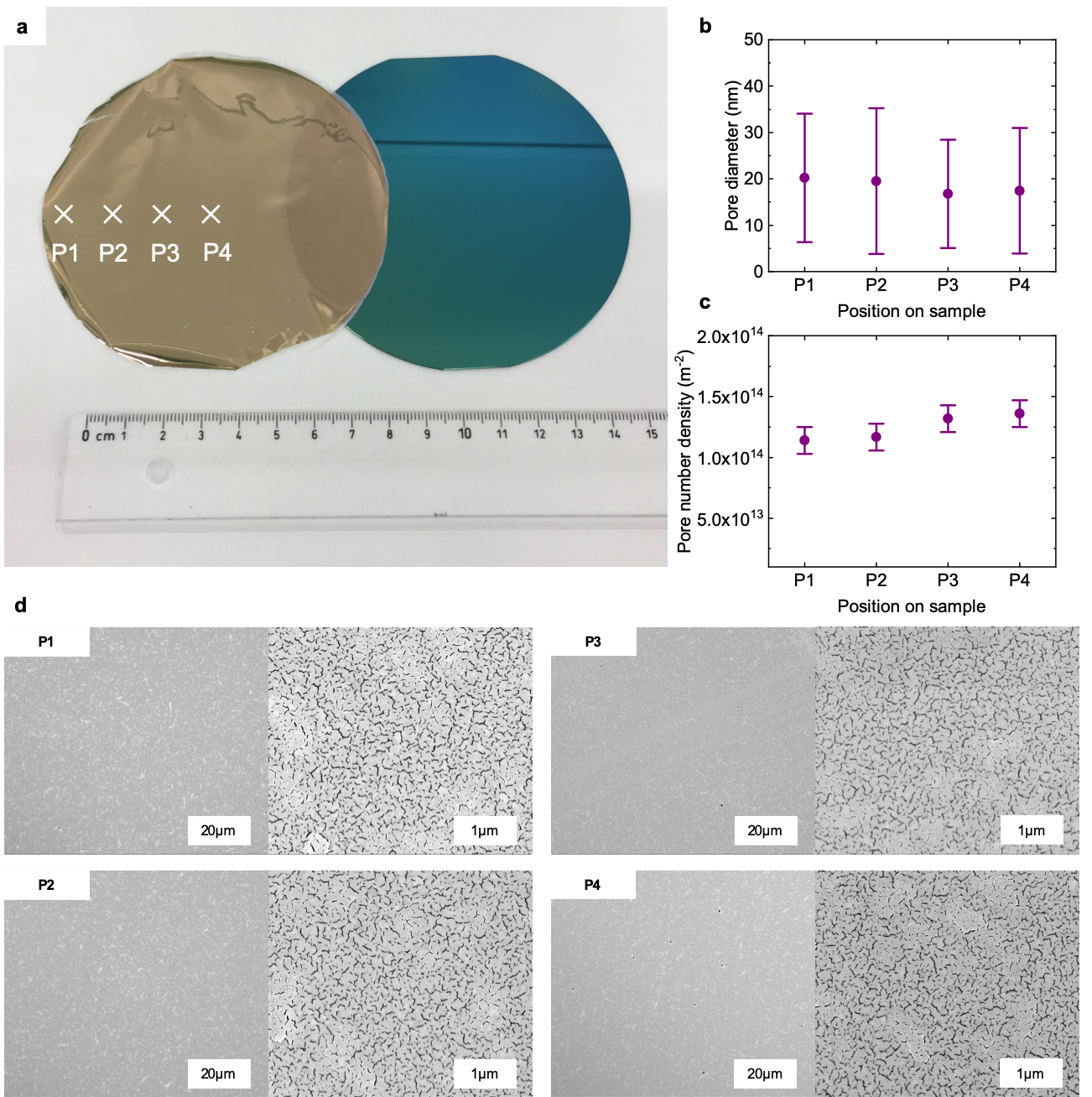}
	\caption{(a) 4-inch porous Au membranes on polycarbonate-track-etched (PCTE) with homogeneous (b) pore diameter and (c) pore number density extracted from four different location P1 to P4. In (d) the pores in the metal membrane are visible as black slits and dots, while the PCTE pores are visible as brighter regions in the background (see higher magnification SEM graphs). The metal was transferred to PCTE following the method developed in Ref.~\cite{choi2018multifunctional}. 
	\label{Fig:SI_Largescalegold}}
\end{figure*}

The pores in the Au membranes of \unit[15]{nm} to \unit[25]{nm} in thickness are slit-like and circular (Fig.~\ref{fullresults}). While they can be longer than \unit[100]{nm}, their minor dimension and radius are within a relatively narrow band. Using ImageJ, we extracted the area of $\sim$ 100 - 150 pores for characterization using an equivalent diameter modelling them as round pores.

Uniform and large-scale porous gold membranes on polycarbonate-track-etched (PCTE) can be formed, demonstrating the ability to yield thin-film porous metal membranes on a polymer support as large as a 4-inch wafer, shown for a \unit[20]{nm}-thick PAuM in Fig.~\ref{Fig:SI_Largescalegold}(a), with a homogeneous  pore diameter and pore number density shown in Figs.~\ref{Fig:SI_Largescalegold}(b) and (c), respectively. 

\section{Gas transport through Au Membranes}
To describe gas flow through our membranes is based on transmission probabilities, and that purely viscous or effusive transport are insufficient to describe our experimental observations. In general, N$_2$ flow through thin membranes with pores in the range $<\unit[50]{nm}$ cannot be described by viscous transport as the mean free path (MFP) at standard temperature and pressure is $\sim \unit[70]{nm}$, where the Knudsen number ($K_n = \mathrm{MFP}/l_{\rm char}$, $l_{\rm char}$ is the characteristic dimension of the channel) is $\sim 1$, indicating the onset of free molecular flow. 

The transport through 2D membranes, e.g. graphene membranes \cite{RN95}, is governed by effusion due to the negligible thickness if the respective pore size is smaller than the MFP. Effusion through a membrane with porosity $\chi$ is given by
\begin{equation}
    J=\frac{\chi}{\sqrt{2\pi M R T}},
    \label{eq:Effusion}
\end{equation}
where $M$ is the molar mass of N$_2$ (\unit[0.028]{kg/mol}), $R$ is the universal gas constant and $T$ is the absolute temperature. It should be noted that in case of purely effusive transport the transport through an ensemble of pores is equivalent to the transport of a single pore with equivalent open area. In a system such as our gold membranes, having non-zero thickness $t$, however, the permeance based on effusion vastly overpredicts the mass transport, c.f. Fig.~2(e) of the main paper, in particular for lower porosities. 

The proper way to describe the gas transport through our porous Au membrane is a combination of Knudsen Diffusion and Effusion~\cite{dushman1922production}. However, the linear superposition as described in~\cite{dushman1922production} was shown to be incorrect, leading to the introduction of transmission probabilities \cite{clausing1932stromung} instead, additionally taking into account the non-circular shape of the majority of the pores. As the pores in our PAuM have aspect ratios 1 $<$ a/b $<$ 10, we calculate the permeance using the extreme cases that all pores are circular (a/b = 1, maximum permeance) and slit-like (a/b = 10, minimum permeance) to compare the corresponding range of values with our experimental data in Fig.~2(e). For membranes with low porosity (\unit[20]{nm}, \unit[25]{nm}), the modelled and measured permeances agree well, while for \unit[15]{nm} membranes, a higher permeance is measured than anticipated by the model. We attributed this behaviour to few larger holes in the membranes or the fact that $t/D$ is small, so the thickness effect is negligible and the flow behavior is mostly effusive.  

\begin{figure}[pt]
	\includegraphics[width=13cm]{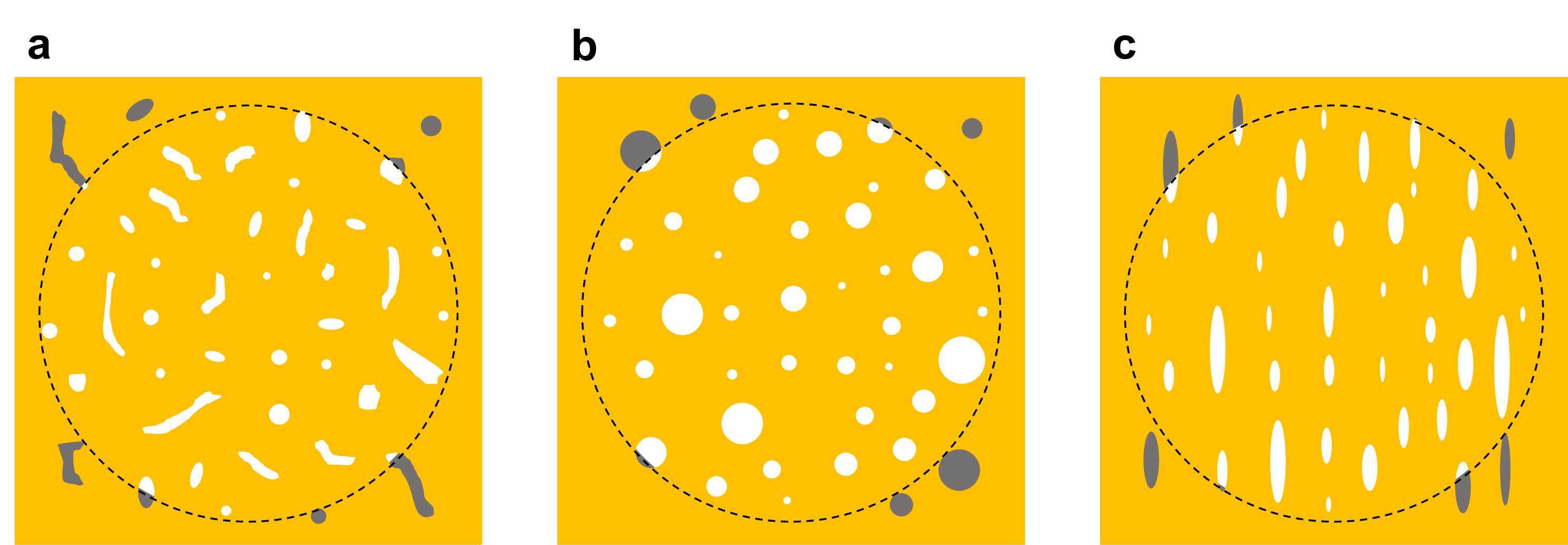}
	\caption{Schematic for modelling the gas transport through Au membranes with (a) realistic pore shapes, (b) circular pores having same areas as in (a) and (c) slit-like pores having same areas as in (a). The dashed black circle outlines the freestanding part of the membrane.  
		\label{Fig:SI_Au_Gas}}
\end{figure}

In this section, we model the gas transport through Au membranes and discuss the effect of the pores' shape on the permeation characteristics of the membrane.  Fig.~\ref{Fig:SI_Au_Gas}(a) schematically depicts a realistic pore shape distribution as it could be found in a membrane. Fig.~\ref{Fig:SI_Au_Gas}(b) depicts the same pore area as in (a), however with circular pores and (c) depicts all pores with a high aspect ratio. Note that (b) and (c) correspond to the upper and lower limit of permeation through a membrane with the corresponding pore size and thickness $t$.

\begin{figure}[h]
	\includegraphics[width=15cm]{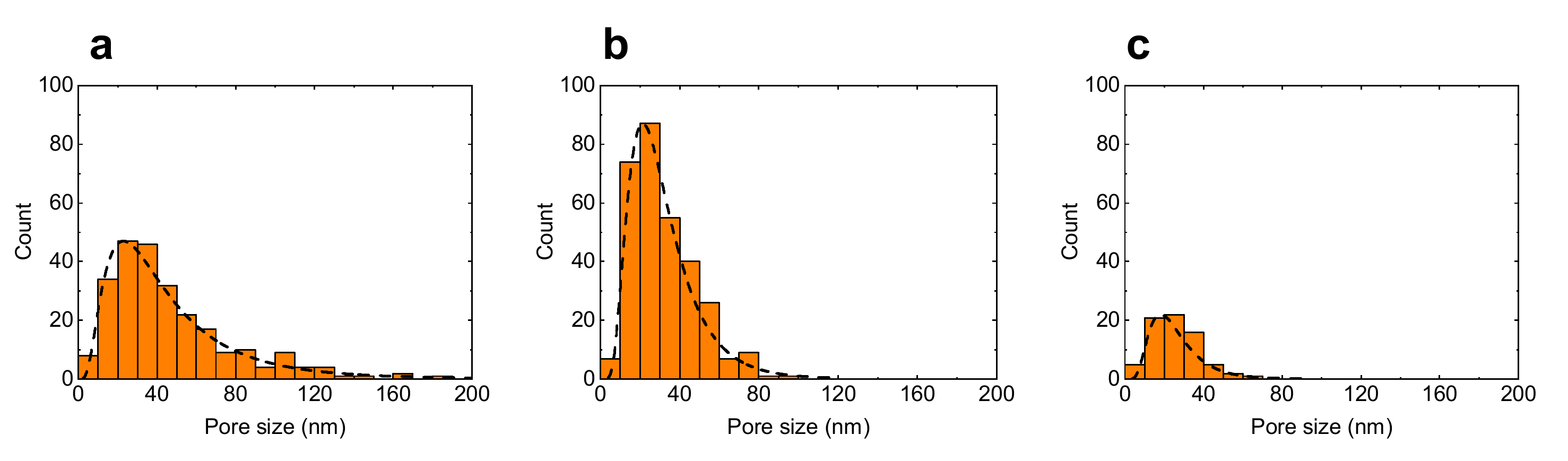}
	\caption{Pore size (= 2$\bar{r}$) histograms of (a) 15nm, (b) 20nm and (c) 25nm Au membranes. The black dashed line marks a corresponding log-normal distribution.
		\label{Fig:SI_Au_Histo}}
\end{figure}

(i) \textbf{Circular pores with depth $t$:} The extracted pore areas from ImageJ were converted into a radius of an equivalent pore. In Fig.~\ref{Fig:SI_Au_Histo}, the histograms of the pore sizes (= 2$\bar{r}$) for the membranes \unit[15]{nm} - \unit[25]{nm} are shown, where the black dashed line marks a corresponding log-normal distribution.

The log-normal distribution is characterized by the distribution function P($m$,$s$), where $m$ and $s$ are the mean and standard deviation of the natural logarithm of the pore radius, respectively. The corresponding pore radius and standard deviation are denoted $\bar{r}$ and $\sigma_{\bar{r}}$, respectively, and are shown in  Table~\ref{table:s_m}. 

\begin{table}[b]
    \centering\begin{tabular}{ l|c|c|c|c|c|c|c|c} 
    Membrane    & $m$ (-) & $s$ (-) & $\bar{r}$ (nm) & ${\sigma_{\bar{r}}}$ (nm) & $t/\bar{r}$ & $t/\bar{r}_{upper}$ & $t/\bar{r}_{lower}$ & W \\ \hline
    \unit[15]{nm} & 2.92     & 0.699   & 23.9                       & 19 & 0.63 & 0.35 & 3.06 & 0.38 - 0.86 \\ 
    \unit[20]{nm} & 2.6       & 0.52     & 16                          & 8.91 & 1.25 & 0.8 & 2.82 & 0.43 - 0.71\\ 
    \unit[25]{nm} & 2.44     & 0.49     & 13.06                     & 6.82 & 1.83 & 1.22 & 3.7 & 0.38 - 0.63 \\ 
\end{tabular}
\caption[Average pore size and standard deviation of Au membranes based on log-normal distribution.]{Average pore size, standard deviation of Au membranes based on log-normal distribution, thickness-to-pore-size-ratio and the corresponding transmission probabilities W.}
\label{table:s_m} 
\end{table}

$t/\bar{r}$ denotes the average ratio of thickness to pore size, with $t/\bar{r}_{upper}$ and $t/\bar{r}_{lower}$ being the upper and lower limit of the ratio based on the average pore size. The corresponding thickness-dependent transmission probabilities $W_t$ are tabulated \cite{clausing1932stromung}, where $Q$ = $Q_e$ * $W_t$, $Q$ is the real permeance and $Q_e$ is the purely effusive permeance. Based on the transmission probabilities and assuming circular pores with depth $t$, one can expect permeances to be reduced up to 62\% of the original value (shown as round dots in Fig.~1e).  The remaining discrepancy can be attributed the effect of pore shape. 

(ii) \textbf{Slit-like pores with depth $t$:} A factor that has been neglected was the pore shape, assuming circular pores previously to allow a more simple modelling. A significant portion of pores in the membrane is, however, non-circular/slit-like. The problem has been studied for long tubes, where at equal area, the gas transport through non-circular pores is always lower compared to a circular pore \cite{steckelmacher1978effect}. For example, the permeance of a slit $Q_e$ with an aspect ratio of a/b = 1/10 is reduced by a shape dependent factor $W_s$ of 0.26 compared to the permeance $Q_c$ circular pore of the same area ($i.e$ $Q_e$ = $W_s$ * $Q_c$). It is intuitive to assume that the flow rate through a short elliptical aperture is therefore reduced compared to a circular aperture, however, the exact factor has yet to be established. The values given in Fig.~2(e) of the main paper therefore show the data assuming round pores. 

\section{Freestanding graphene as Raman reference}

\begin{figure}[hb]
	\includegraphics[width=8.7cm]{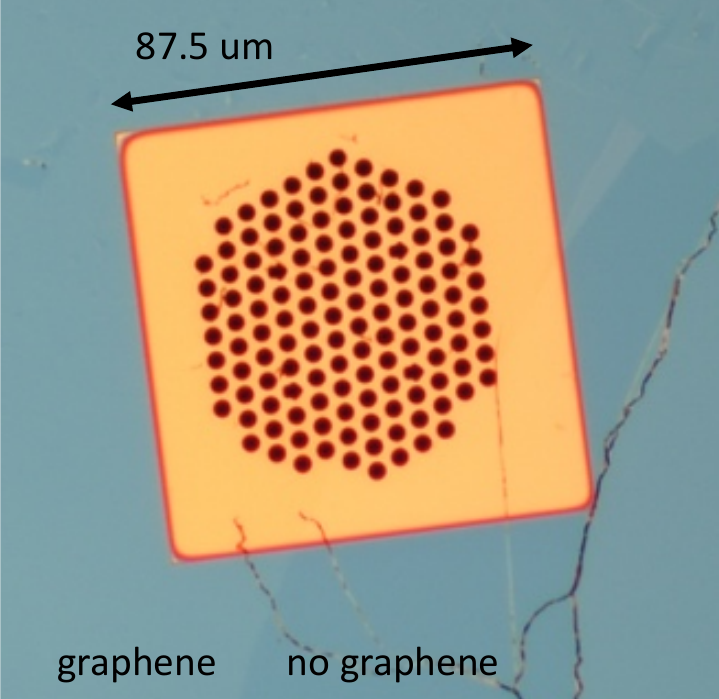}
	\caption{Mechanically exfoliated graphene suspended over a patterned SiN membrane were used as Raman reference. The contrast outside the patterned area indicated that most of the holes are covered by graphene.  
		\label{Fig:SI_SH_FigA}}
\end{figure}

Natural graphite was mechanically exfoliated onto an oxidized silicon wafer. A monolayer was identified optically, see \cite{blake2007making}, and then transferred to a patterned silicon nitride (SiN) membrane shown in Fig.~\ref{Fig:SI_SH_FigA} following the procedure outlined in Ref.~\cite{reina2008transferring}. Reference Raman spectra were obtained by focusing the laser centrally on the monolayer graphene suspended of the circular holes in the SiN membrane. The SERS Raman spectra in Fig.~3 of the main paper were normalized to the $2D$-intensity (integrated area) of suspended graphene. To obtain the enhancement factors given in Table~1 of the main paper, we normalized the SERS intensities of the $G$- and $2D$-modes individually to the corresponding intensities of the suspended graphene Raman reference for each excitation wavelength.

\clearpage
\section{SERS of two additional graphene covered, suspended Au-membranes}

\begin{figure*}[h]
	\includegraphics[width=\textwidth]{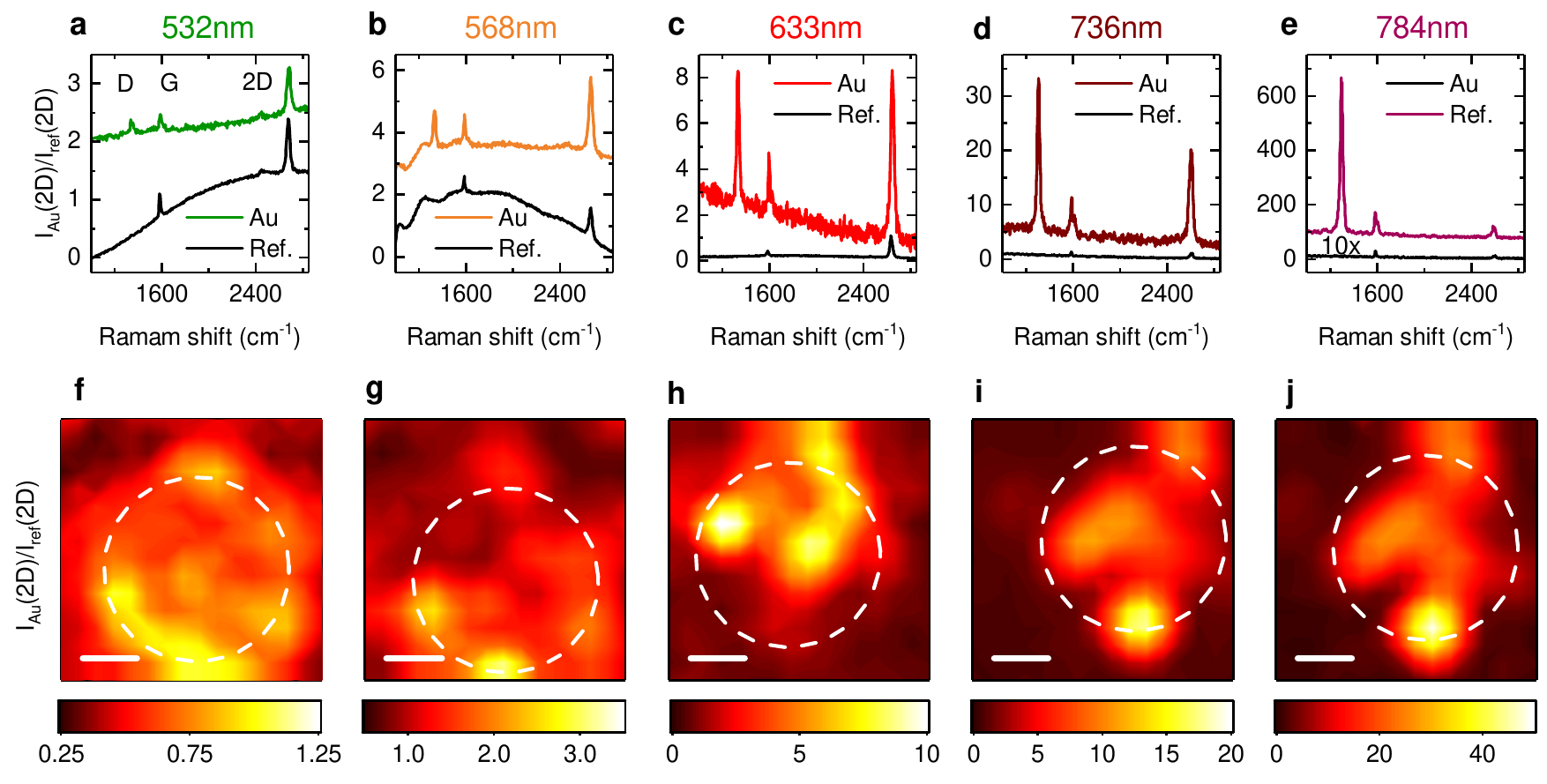}
	\caption{$2D$-mode SERS enhancement map of graphene placed on freestanding Au membrane AuM-R1 for five  excitation wavelength. The intensity is normalized to the $2D$-mode of freestanding graphene. (a-e) Raman spectra(red) extracted from the Raman maps (f-j) at the location of the highest $2D$ enhancement together with reference spectra (black) of freestanding graphene. The scale bar in (f-j) is $500\,$nm.  
		\label{Fig:SI_SH_FigB}}
\end{figure*}

\begin{figure*}[h]
	\includegraphics[width=\textwidth]{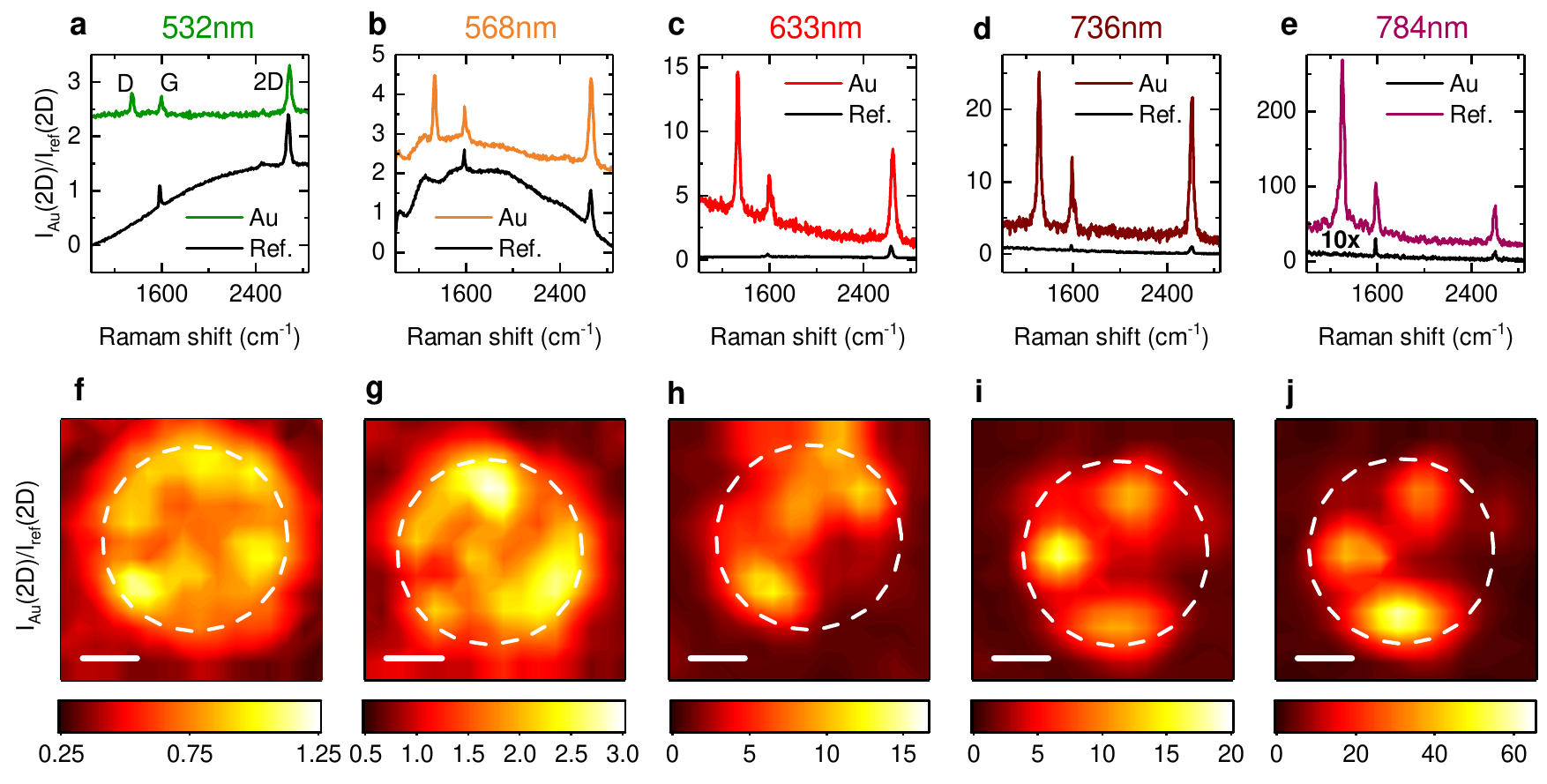}
	\caption{$2D$-mode SERS enhancement map of graphene placed on freestanding Au membrane AuM-R2 for five  excitation wavelength. The intensity is normalized to the $2D$-mode of freestanding graphene. (a-e) Raman spectra(red) extracted from the Raman maps (f-j) at the location of the highest $2D$ enhancement together with reference spectra (black) of freestanding graphene. The scale bar in (f-j) is $500\,$nm.\label{Fig:SI_SH_FigC}}
\end{figure*}

\begin{figure*}[t]
	\centering
	\includegraphics[width=\textwidth]{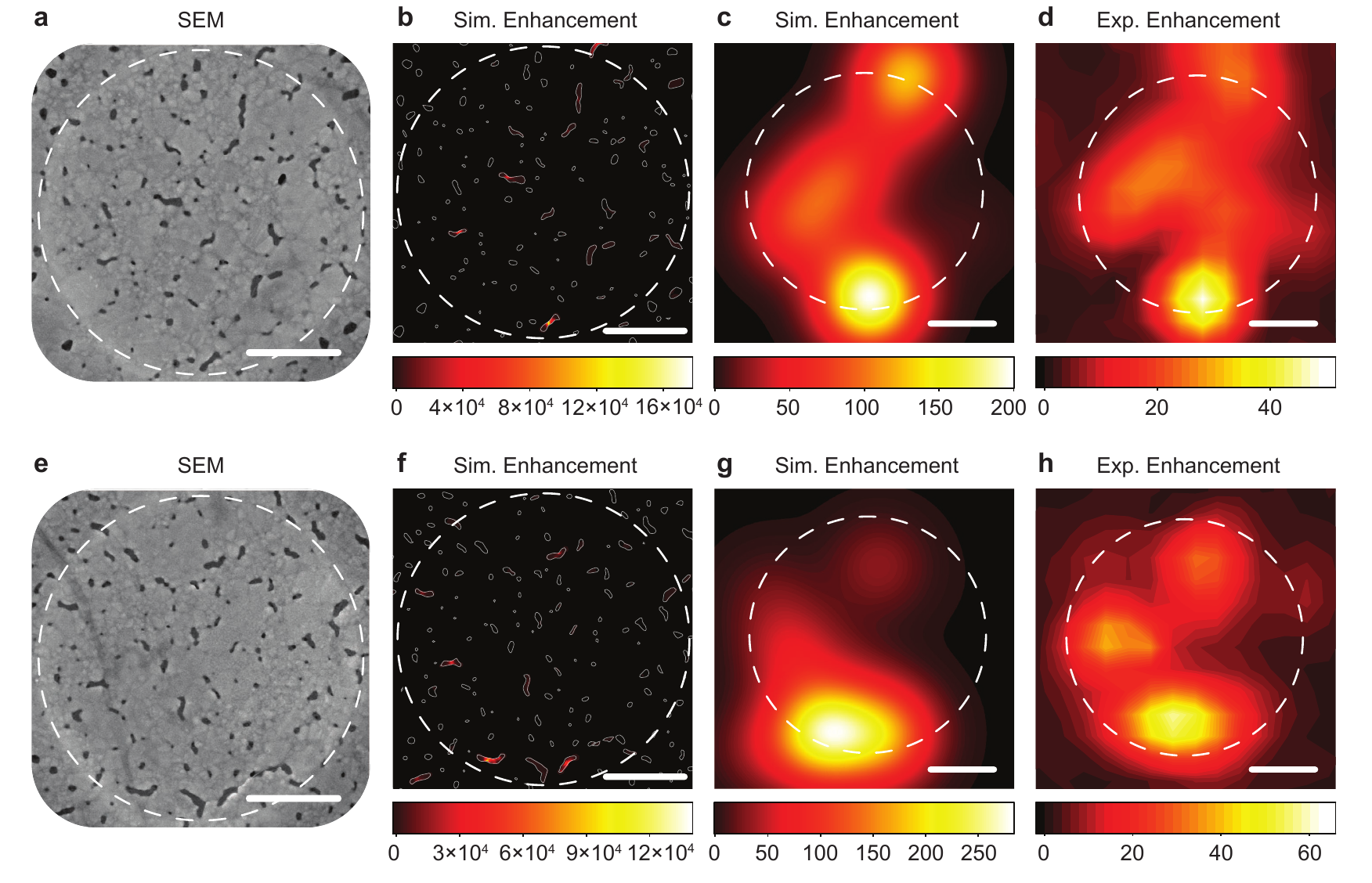}
	\caption{Numerical analysis of two additional porous Au membranes (PAuM-R1/R2, see Fig.~\ref{Fig:SI_SH_FigC}). (a,e) SEM images of a freestanding Au membranes (\unit[25]{nm} nominal thickness), covered by monolayer graphene. (b,f)~Simulated local enhancement $\mathrm{EF}_{\rm loc}$ (cf. main text) of the 2D Raman mode intensity for an excitation wavelength $\lambda_{\rm exc}$ of \unit[784]{nm}, overlayed with the outlines extracted from (a,e). (c,g) As (b,f), convoluted with a Gaussian function, the full width at half maximum (FWHM) of which is given by $\lambda_{\rm exc} / (2 \mathrm{NA})$, where $\mathrm{NA}=0.8$ is the numerical aperture of the objective. (d,h)~Experimental map of the enhancement of the integrated 2D intensity at $\lambda_{\rm exc} = \unit[784]{nm}$, reproducing the data shown in Fig.~\ref{Fig:SI_SH_FigB}(j) and Fig.~\ref{Fig:SI_SH_FigC}(j), respectively. The dashed outlines mark the position of the suspended membrane. Scale bar is \unit[500]{nm}.
		\label{MPfigA}}
\end{figure*}

In this section we present the SERS enhancement of two additional graphene-covered, suspended Au-membranes shown Figs.~\ref{Fig:SI_SH_FigB} and .~\ref{Fig:SI_SH_FigC}, respectively. The experimentally observed SERS enhancement is summarized in Table~\ref{tab:SERS_enhancments}, while the comparison to numerical simulations for both membranes is shown in Fig.~\ref{MPfigA}.

\clearpage

\begin{table*}[t]
    \centering
    \begin{tabular}{c|c|c|c|c|c|c}
     $\lambda$(nm)& 532 & 568 & 633 & 736 & 784 \\
     \hline
     EF$_G$ & 0.9 & 3.0 & 13.3 & 169 & 202 & main paper\\
     EF$_2D$ & 1.0& 4.3 & 10.5 & 42.7 & 60.0 & main paper\\
     \hline 
     EF$_G$ & 1.3 & 2.2 & 20.3 & 30.5 & 62.8 & AuM-R1\\
     EF$_2D$ & 1.1 &  2.90 & 12.0   & 24.0 & 54.8 & AuM-R1\\
     \hline 
     EF$_G$ & 1.0 & 2.2 & 8.8 & 18.7 & 56.6 & AuM-R2\\
     EF$_2D$ & 1.0 & 3 & 9.7   & 22.3 & 46.1 & AuM-R3\\
    \end{tabular}
    \caption{Summary of SERS enhancement factors of the G- and 2D mode at locations of highest intensity for the freestanding PAuM in the main paper, and the two additional PAuM (PAuM-R1 see Fig.~\ref{Fig:SI_SH_FigB} , PAuM-R2 see Fig.~\ref{Fig:SI_SH_FigC}) for different excitation wavelengths.}
    \label{tab:SERS_enhancments}
\end{table*}
\clearpage
\section{Numerical analysis of regular slot antennas in a thin Au film.}
\begin{figure*}[h]
	\centering
	\includegraphics[width=\textwidth]{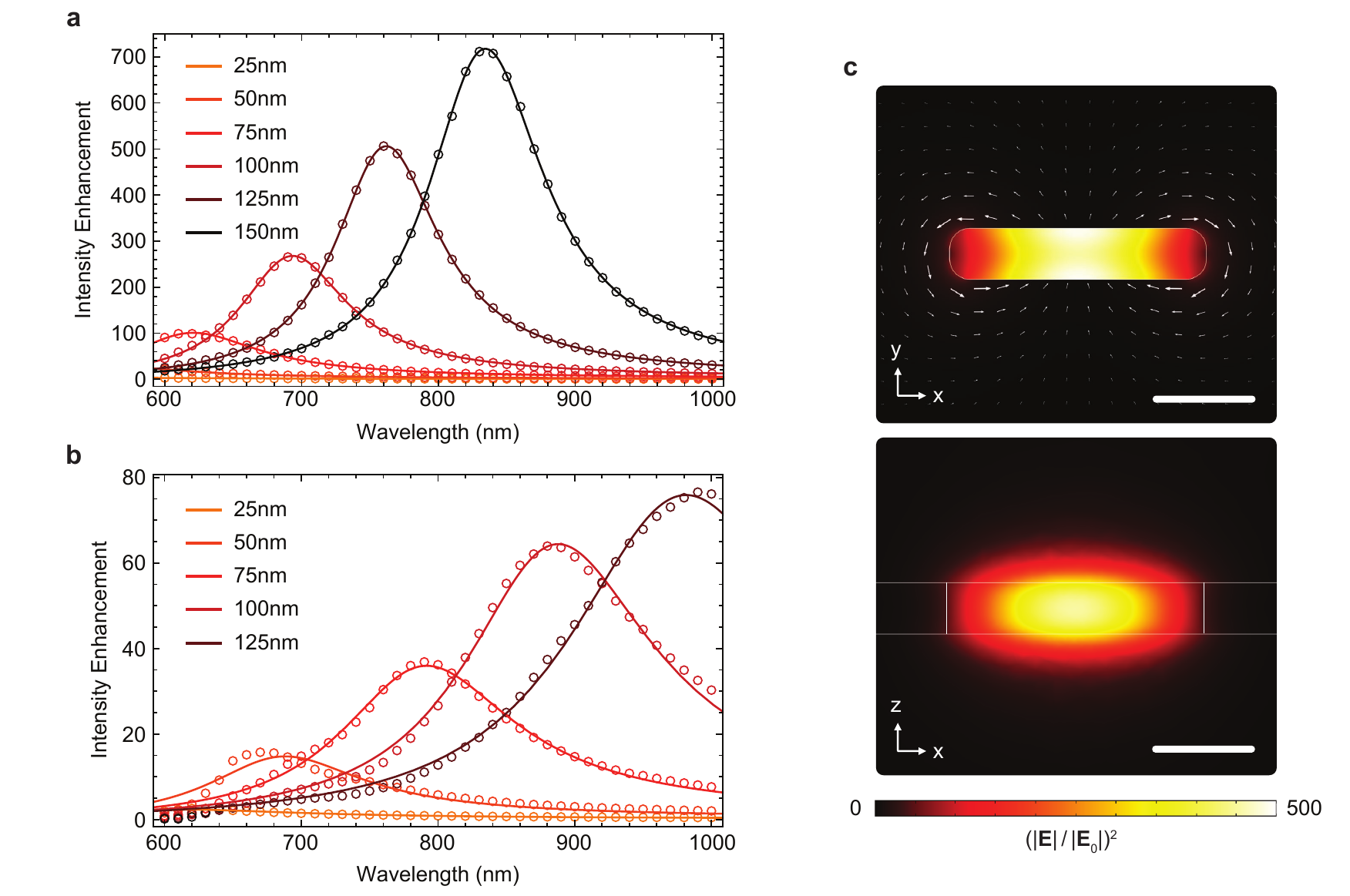}
	\caption{Numerical analysis of regular slot antennas in a thin Au film. (a) Simulated intensity enhancement $|\mathbf{E}|^2/|\mathbf{E}_0|^2$ in the geometrical center of rectangular slots of constant width $w=\unit[25]{nm}$ and height $h=\unit[25]{nm}$, a corner radius of $\unit[10]{nm}$ and varying length ranging from \unit[25]{nm} to \unit[150]{nm} in steps of \unit[25]{nm}. The slots are suspended (surrounded by air) and illuminated with plane waves, polarized parallel to their short axis. (b) As (a) but for supported slot antennas (Si$_3$N$_4$ substrate with a refractive index of $n=2$). Simulations are carried out as a function of wavelength (circles), fitted by Lorentzian functions (lines). (c) Cross sectional intensity enhancement distribution of a $\unit[125 \times 25 \times 25]{nm^3}$ suspended slot antenna within a $x$-$y$-plane (top) and a $x$-$z$-plane (bottom) through the center of the slot antenna. White arrows indicate magnitude and direction of the current density. Scale bar is \unit[50]{nm}.
		\label{MPfigB}}
\end{figure*}
\clearpage

\section{SERS stability under high illumination powers}
\begin{figure*}[h]
	\centering
	\includegraphics[width=8.7cm]{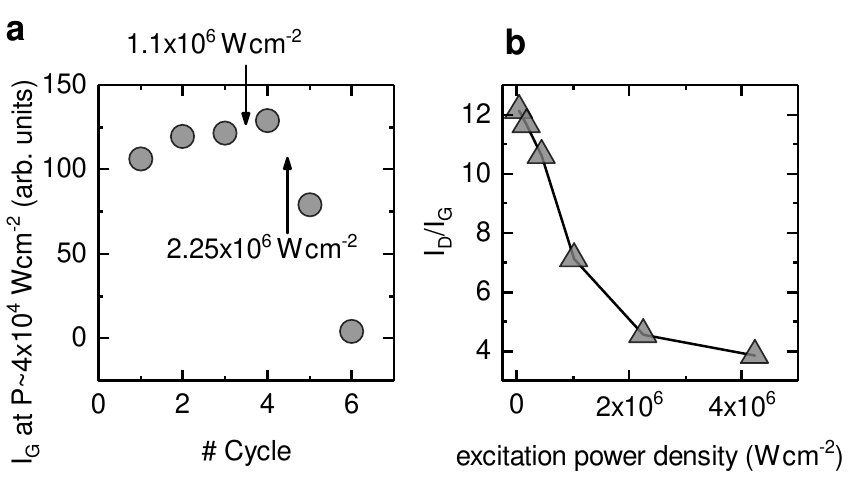}
	\caption{(a) G-mode intensity at nanopore SERS hotspot measured with low excitation power densites P after every increase in P. For $\mathrm{P_{min}}>\unit[1.1\times10^6]{W cm^{-2}}$, the decrease of the the G-mode intensity indicates the onset of irreversiable damage to the graphene at the SERS hotspot. (b) $\mathrm{I_D/I_G}$ ratio as a function of P, demonstrating the restoration of defects at the SERS hotspot with increasing temperature.
		\label{Fig:SI_SH_Cycling}}
\end{figure*}

In Fig.~\ref{Fig:SI_SH_Cycling}(a) we present the G-mode intensity measured at $\mathrm{P_{min}}\sim\unit[4\times10^4]{Wcm^{-2}}$ after every increase in excitation power density P, which serves as a control measurement to ensure that the graphene/nanopore system forming the SERS hotspot is stable with every increase in illumination power. The G-mode intensity remains approximately constant up to $\mathrm{P} = \unit[1.1\times10^6]{W cm^{-2}}$, and drops for $\mathrm{P}>\unit[2.25\times10^6]{W cm^{-2}}$, which indicates the failure of the graphene/nanopore hotspot, see main paper.

We cannot use the graphene D-mode, which indicates the presence of defects, to monitor the onset of damage with increasing P because the nonzero intrinsic defect density in the CVD graphene used to probe the nanopore hotspot. In fact, the $\mathrm{I_D/I_G}$ ratio which typically serves as an indicator for the defect density, counterintuitively decreases with increasing excitation power density as shown in Fig.~\ref{Fig:SI_SH_Cycling}. We explain this behaviour by the increase in temperature at the hotspot with increasing excitation power, c.f. main paper, as certain types of defects in graphene have low restoration energies such that defects in the graphene membrane can effectively be repaired at moderate temperatures~\cite{botari2016graphene}. 

In Table~\ref{tab:SERS_powerdensities}, we list a representative set of SERS structure including the powerthreshold to avoid damage. The PAuM introduced in this work exceed all other structures by 2-3 orders of magnitude. 

\begin{table*}[t]
    \footnotesize
    \centering
    \begin{tabular}{l|l|l|l|l|l}
     Structure & fabrication  & probe & $\lambda$ [nm]  & power [W/cm$^{-2}$]  & Ref.\\
     \hline
    Au dimers with sub-nm gaps & e-beam & molecules &\unit[700 - 900]{} & \unit[$2.2\times 10^3$]{} &  \citenum{zhu2014quantum} \\
      Silver nanodiscs & e-beam & molecules & \unit[633]{} & \unit[$5.5\times 10^4$]{} & \citenum{RN964}\\
       Au Nanotriangles & self-assembly & molecules & \unit[633\,\&\,785]{} & \unit[$1.7-6.9\times 10^3$]{} & \citenum{RN963}\\
     "Sea urchin" Au mesoparticles &  self-assembly & molecules & \unit[633\,\&\,785]{} & \unit[$1.4-6.4\times 10^3$]{} & \citenum{RN962}\\
    Au nanodot arrays & mask lithography & molecules & \unit[633]{} & \unit[1.2]{} & \citenum{RN958}\\
     Au dimers & e-ebam & graphene & \unit[633]{} & \unit[$5\times 10^4$]{} & \citenum{mueller2017evaluating}\\
      Au dimers & e-ebam & graphene & \unit[633]{} & \unit[$5\times 10^4$]{} & \citenum{wasserroth2018graphene}\\
      \hline
      PAuM & evaporation & graphene & \unit[532-785]{} & \unit[$1.1\times 10^6$]{} & this work\\
      \hline
    \end{tabular}
    \caption{Comparison of maximum power densities used for a variety of SERS structures and the porous Au membranes introduced in this work.}
    \label{tab:SERS_powerdensities}
\end{table*}

\section{Large-scale SERS mapping of 20 nm thick porous Au membrane}
\begin{figure*}[t]
	\centering
	\includegraphics[width=12cm]{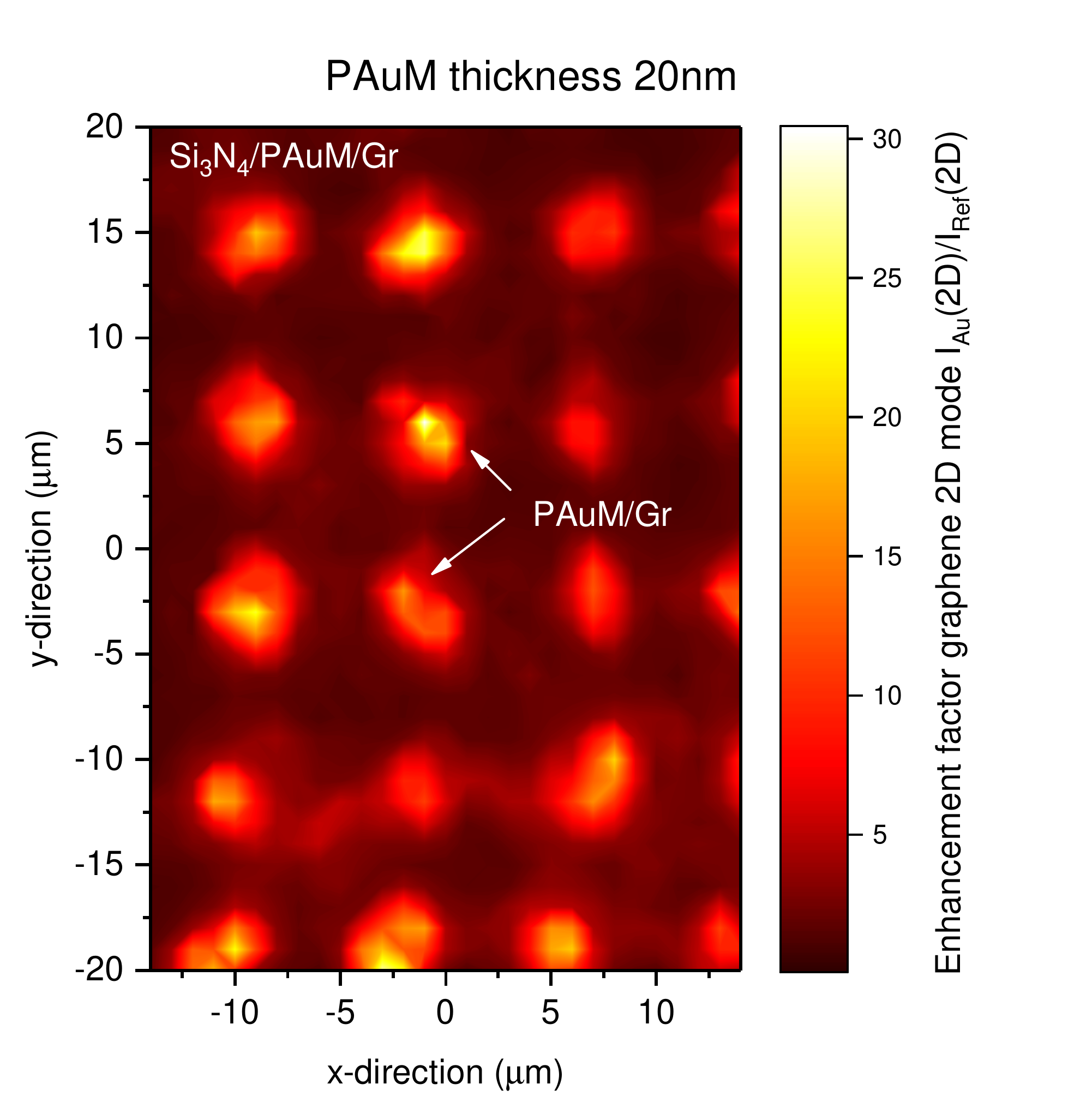}
	\caption{Large-scale SERS map of the 2D-mode of graphene placed on top of a $\unit[20]{nm}$ PAuM for an excitation wavelength $\unit[633]{nm}$. As in the main paper, the enhancement is given using freely suspended graphene as reference. SERS enhancement occurs predominantly at the circular, freestanding PAuM regions, and is barely present in the areas supported by the Si$_3$N$_4$ frame.}
		\label{Fig:SI_SH_PAuM_20nm}
\end{figure*}

Figure~\ref{Fig:SI_SH_PAuM_20nm} shows a large scale 2D-mode SERS map of $15$ freestanding PAuM segments covered with graphene for $\unit[633]{nm}$ excitation. Taking into account the locations of maximium enhancement of each circular segment, we obtain an average 2D-mode enhancement of $19\pm6$, which is a factor of $1.73$ larger than for the three freestanding segments of the  $\unit[25]{nm}$ thick PAuM ($11\pm1$, see main paper). The increase in SERS enhancement is in good agreement with the corresponding $1.9$-fold increase in pore density from $\unit[4.4\times10^{13}]{m^{-2}}$ for the $\unit[25]{nm}$ membrane to $\unit[8.5\times10^{13}]{m^{-2}}$ for the $\unit[20]{nm}$, see Figure~2 of the main paper. We conclude that the total SERS enhancement increases approximately linearly with the pore density number of the porous Au membranes for thicknesses of \unit[15-25]{nm} as investigated here.

\end{document}